\documentclass[aps,prb,twocolumn,amsmath,amssymb,superscriptaddress,floatfix,reprint,longbibliography]{revtex4-2}
\usepackage[latin1]{inputenc}
\usepackage{amsfonts}
\usepackage{lipsum}
\usepackage{bbold}
\usepackage{hyperref}
\usepackage{graphicx}
\usepackage[usenames,dvipsnames]{color}
\usepackage{float}
\usepackage{placeins}
\usepackage{multirow}
\usepackage{tabularx} 
\usepackage{dcolumn}
\usepackage{soul}
\usepackage{physics}
\usepackage{braket}
\usepackage{mathtools}
\usepackage[range-units=single,separate-uncertainty]{siunitx}
\usepackage{verbatim} 
\usepackage{layouts}

\begin{document}

\author{Philipp Eck}
\affiliation{Institut f\"ur Theoretische Physik und Astrophysik, Universit\"at W\"urzburg, D-97074 W\"urzburg, Germany}
\affiliation{W\"urzburg-Dresden Cluster of Excellence ct.qmat, Universit\"at W\"urzburg, D-97074 W\"urzburg, Germany}

\author{{Carmine Ortix}}
\affiliation{Dipartimento di Fisica ``E. R. Caianiello", Universit\`a di Salerno, IT-84084 Fisciano (SA), Italy}

\author{Armando Consiglio}
\affiliation{Institut f\"ur Theoretische Physik und Astrophysik, Universit\"at W\"urzburg, D-97074 W\"urzburg, Germany}
\affiliation{W\"urzburg-Dresden Cluster of Excellence ct.qmat, Universit\"at W\"urzburg, D-97074 W\"urzburg, Germany}
 
\author{Jonas Erhardt}
\affiliation{Physikalisches Institut, Universit\"at W\"urzburg, D-97074 W\"urzburg, Germany}
\affiliation{W\"urzburg-Dresden Cluster of Excellence ct.qmat, Universit\"at W\"urzburg, D-97074 W\"urzburg, Germany}

\author{Maximilian Bauernfeind}
\affiliation{Physikalisches Institut, Universit\"at W\"urzburg, D-97074 W\"urzburg, Germany}
\affiliation{W\"urzburg-Dresden Cluster of Excellence ct.qmat, Universit\"at W\"urzburg, D-97074 W\"urzburg, Germany}

\author{Simon Moser}
\affiliation{Physikalisches Institut, Universit\"at W\"urzburg, D-97074 W\"urzburg, Germany}
\affiliation{W\"urzburg-Dresden Cluster of Excellence ct.qmat, Universit\"at W\"urzburg, D-97074 W\"urzburg, Germany}


\author{Ralph Claessen}
\affiliation{Physikalisches Institut, Universit\"at W\"urzburg, D-97074 W\"urzburg, Germany}
\affiliation{W\"urzburg-Dresden Cluster of Excellence ct.qmat, Universit\"at W\"urzburg, D-97074 W\"urzburg, Germany}

\author{Domenico Di Sante}
\affiliation{Department of Physics and Astronomy, University of Bologna, 40127 Bologna, Italy}
\affiliation{Center for Computational Quantum Physics, Flatiron Institute, New York, 10010 NY, USA}

\author{Giorgio Sangiovanni}
\email{e-mail: sangiovanni@physik.uni-wuerzburg.de}
\affiliation{Institut f\"ur Theoretische Physik und Astrophysik, Universit\"at W\"urzburg, D-97074 W\"urzburg, Germany}
\affiliation{W\"urzburg-Dresden Cluster of Excellence ct.qmat, Universit\"at W\"urzburg, D-97074 W\"urzburg, Germany}


\title{Real-space Obstruction in Quantum Spin Hall Insulators}

\date{\today}
\begin{abstract}
The recently introduced classification of two-dimensional insulators in terms of topological crystalline invariants has been applied so far to ``obstructed" atomic insulators characterized by a mismatch between the centers of the electronic Wannier functions and the ionic positions.
We extend this notion to quantum spin Hall insulators in which the ground state cannot be described in terms of time-reversal symmetric localized Wannier functions.
A system equivalent to graphene in all its relevant electronic and topological properties except for a real-space obstruction is identified and studied via symmetry analysis as well as with density functional theory. 
The low-energy model comprises a local spin-orbit coupling and a non-local symmetry breaking potential, which turn out to be the essential ingredients for an obstructed quantum spin Hall insulator.
An experimental fingerprint of the obstruction is then measured in a large-gap triangular quantum spin Hall material. 

\end{abstract}
\maketitle

\section*{Introduction}
\label{sec:Introduction}


Insulating phases of matter are categorized based on topological properties of their band structures~\cite{kane2005quantum,kane2005z,Fu2007topological,Fu2007topologicainversion,hasan2010colloquium}. In trivial insulators
the valence bands are adiabatically connected to an atomic limit and exponentially localized Wannier functions must exist~\cite{brouder2007exponential,panati2007triviality}.
Hence, in the absence of a Wannier representation, the corresponding phase is topologically non-trivial as, for instance, the case of quantum spin Hall insulators (QSHIs) in the presence of time-reversal symmetry~\cite{fu2006time,roy2009z2,Loring2010disordered,Soluyanov2011wannier,Soluyanov2012smooth}.
The bulk-boundary correspondence guarantees that QSHIs possess gapless anomalous boundary modes, whose spin-momentum locking and topological robustness are particularly attractive for spintronics applications. 
A subset of all possible non-trivial phases can be identified by means of the topological quantum chemistry (TQC) approach. This relies on symmetry-indicators, that reveal the existence or lack of elementary band representations~\cite{Po2016filling,bradlyn2017topological,kruthhof2017topological,po2017symmetry,cano2018building}.

Recently, a further real-space classification has been put forward for trivial insulators. 
Depending on the spatial localization of the Wannier centers for the valence bands  with respect to the underlying ions' lattice positions, one distinguishes between conventional and ``obstructed'' atomic insulators~\cite{Po2016filling,bradlyn2017topological,po2017symmetry,cano2018building,obstructed1,obstructed2,obstructed3}.
In the latter ones, at least one of the Wannier functions is displaced away from the lattice site positions, leading to interesting effects in open boundary condition geometries, 
such as metallic interface states or higher-order topological phases in 2D~\cite{benalcazar2017quantized,benalcazar2017electric,langbehn2017reflection,song2017d,schindler2018higher,schindler2018higher2,Miert2018higher,Miert2020topological,eck2022recipe}.
Provided that, in a specific case, symmetry indicators do exist, obstructed atomic insulators can then be unambiguously determined with TQC, at odds with the identification of non-trivial $\mathbb Z_2$-phases where TQC can predict false negatives~\cite{bradlyn2017topological,po2017symmetry}. 

%

Extending the concept of real-space obstruction to QSHIs would constitute an intriguing step. This is conceptually possible since symmetry-protected topological phases lack a Wannier representation only when using Wannier function basis set that preserve the protecting internal and/or point-group symmetries. A description of a QSHI breaking the requirement that Wannier functions must come in Kramers' degenerate pairs is therefore completely allowed. This has been in fact successfully achieved by Soluyanov and Vanderbilt in Ref.~\cite{Soluyanov2011wannier}. Note that the symmetry character of the electronic band cannot diagnose the charge centers of such Wannier functions. The Kramers' degeneracy of the electronic bands cannot be removed while preserving the time-reversal symmetry of the full many-body ground state. Therefore, the construction of physical elementary band representations at the basis of TQC always reduces to Wannier Kramers' pairs.

The localized Wannier functions for a QSHI insulator originally introduced for the Kane-Mele model in Ref.~\cite{Soluyanov2011wannier} are centered on the two inequivalent sublattices of the honeycomb net. As we show here, a graphene-like band structure can be also realized by chiral orbitals on a triangular lattice. It turns out that the local degrees of freedom play the role of the sublattice isospin of the  Kane-Mele model for single $p_z$- or $s$-orbitals.
We find that, despite the existence of a formal mapping between these two realizations of Dirac fermions at the valley momenta, the triangular QSHI can be topologically distinct from graphene.
The difference is precisely the real-space obstruction as Wannier functions for the valence bands can be localized away from the triangular atomic positions. The Wannier centers form, in turn, a honeycomb motif as they localize in the void spaces between the atoms, as we show by applying the Soluyanov-Vanderbilt construction to our triangular QSHI.

As we elaborate below, there is an additional physical property distinguishing the two models. This regards the locality of the SOC term that is of second-nearest-neighbor nature in graphene but local in the triangular multi-orbital QSHI.
This has a profound impact not only on the size of the gap in material realizations of our model but also on the real-space obstruction.

In this work we hence discuss a mapping between the valence states of a honeycomb and triangular lattice and show how this relies on the fact that a certain wave function symmetry may be represented by basis sets being localized on different Wyckoff positions. 
Such mapping is however a momentum-dependent concept and by determining the associated absence or presence of a real-space obstruction, we elevate its significance to the entire BZ.
We develop this idea for hexagonal and trigonal space groups, for which we show the equivalence of the Dirac fermions at the valley momenta K/K$^\prime$. By embedding them in a simple tight-binding model in the whole BZ, we unveil the real-space obstruction that emerges in the triangular case only. 
Further, we give full support to our theoretical findings by means of scanning tunneling microscopy (STM) and angle-resolved photoelectron spectroscopy (ARPES) measurements on a recently synthesized material~\cite{bauernfeind2021design,erhardt2022indium}. In the final part, we elaborate on the local orbital angular momentum (OAM) polarization, a decisive difference between the Kane-Mele model and the triangular QSHI besides the real-space obstruction, which also propagates into the edge states.

\section{Equivalent Bloch wave function representations}
\label{sec:emergent_hc}

\begin{figure}
    \centering
    \includegraphics[width=1.0\linewidth]{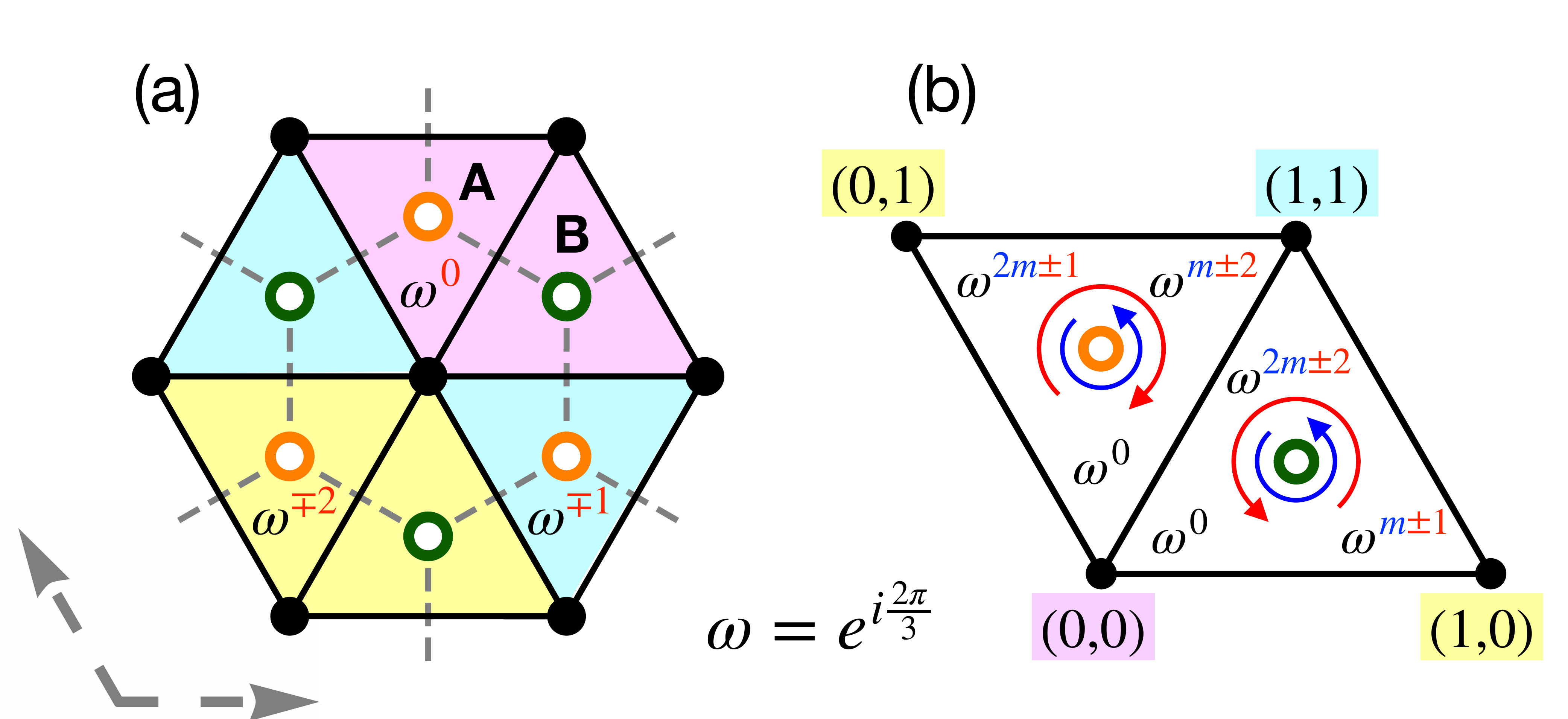}
    \caption{
    (a) Wyckoff positions in the hexagonal unit cell giving rise to the triangular and honeycomb lattices. The former is made of the filled black circle at 1$a$, while the latter by the orange ``A'' and green ``B''  open circles at 2$b$. 
    A different Bloch phase $\operatorname{exp}({i {\bf k}\cdot {\bf R}})$ (purple, yellow and cyan colors) calculated at ${\bf k}$=K or K$^\prime$ is assigned to each unit cell (diamond shapes). The filled black circle in the center corresponds to ${\bf R}=(0,0)$ and represents the atom assigned to the purple unit cell).
     (b) Interference between orbital and Bloch phases. The blue arrow shows how the  phase of the chiral orbitals sitting at the ${\bf R}$ positions of the triangular lattice winds going around the A/B points. In red we denote instead the contribution from the Bloch phases along the same path. For this specific example, we have chosen an orbital with $m\;\text{mod}\;3=1$ at K.}
    \label{fig:model_new}
\end{figure}

The honeycomb and the triangular lattices share the same wallpaper group $p6m$.
The former is bipartite with orbitals located at the Wyckoff positions $2b=\{\mathrm{A}=(1/3,2/3),\mathrm{B}=(2/3,1/3)\}$ \footnote{2$b$ is the notation for wallpaper groups~\cite{benalcazar2019quantization}. For the corresponding three-dimensional space group 191 these positions are denoted by $2c=\{\mathrm{A}=(1/3,2/3,0),\mathrm{B}=(2/3,1/3,0)\}$.}, while the site of the triangular lattice is positioned at $1a=(0,0)$~\cite{aroyo2006bilbao1}.
Here, we will be focusing on the following comparison: the 2$b$ basis site of the honeycomb with A and B sublattices [orange and green open circles in Fig.~\ref{fig:model_new}(a)] on the one hand and, on the other hand, the triangular lattice with its single site (filled black circle) in the unit cell. 
The former has one orbital with magnetic quantum number $m_{hc}=0$ per site while the latter is equipped with two chiral orbitals with $m_{t}=\pm m$ on each atom.
The different dimension of the orbital (flavor) subspaces ensures the same number of degrees of freedom between the two situations.

As we are going to elaborate in this paper, these two lattices possess different electronic band structures but their Bloch eigenstates become indistinguishable at two special points in the BZ: the valley momenta K $=(1/3,1/3)$ and K$^\prime=(-1/3,-1/3)$.
Moreover, the charge density profile from the eigenvectors at K and K$^\prime$ prescinds from the basis. In fact, regardless of having single orbitals at the $2b$ positions or two chiral flavors located at the $1a$ site, the valley charge arranges spatially to form a honeycomb connectivity on the $2b$ Wyckoff positions A and B. 
Such a pattern does not come as a surprise in the case of the honeycomb lattice where the atoms are indeed occupying the positions where the charge accumulates. On the contrary, it is less obvious to see how chiral orbitals on a triangular lattice can form an electronic motif centered on the voids, i.e. on the orange and green open circles in the two triangles of Fig.~\ref{fig:model_new}(b). This may be suggestive of a real-space obstruction in the triangular lattice, which we will more formally elaborate on in the next sections.

\renewcommand{\arraystretch}{1.3}
\newcommand\wE{0.2\textwidth}
\newcommand\ws{0.05\textwidth}
\newcommand\ww{0.44\textwidth}
\newcommand\om[2]{\multicolumn{1}{D{,}{}{-1}|}{{#1},\omega^{#2}}}
\newcommand\sign[2]{\multicolumn{1}{D{,}{}{-1}|}{#1,#2}}
\newcolumntype{x}[1]{>{\centering\arraybackslash\hspace{0pt}}p{#1}}
\begin{table*}[]
    \centering
    \begin{tabular*}{\textwidth}{x{\wE}|x{\ws}|x{\ws}|x{\ws}|x{\ws}|x{\ws}|x{\ws}|x{\ww}}
        \multicolumn{8}{c}{Two-dimensional representations of $D_{3h}$} \\ \hline\hline
                   representation &  $I$ & $2C_3$ & $3C_2^\prime$ & $\sigma_h$ & $2S_3$ & $3\sigma_v$ & orbital basis \\ \hline
        $E^\prime$ &    2 &   $-1$ &             0 &       \sign{}{2} &   \sign{-}{1} &           0 & $(p_x,p_y)$, $(d_{x^2-y^2},d_{xy})$, $(f_{xz^2}, f_{yz^2})$\\[4pt]
$E^{\prime\prime}$ &    2 &   $-1$ &             0 &      \sign{-}{2} &   \sign{}{1} &           0 & $(d_{xz},d_{yz})$, $(f_{xyz}, f_{z(x2-y2)})$\\
        \multicolumn{8}{c}{} \\
        \multicolumn{8}{c}{Chiral one-dimensional representations of $C_{3h}$} \\ \hline\hline
                   representation &  $I$ &     $C_3^1$ &      $C_3^2$ & $\sigma_h$ &     $S_3^1$ &     $S_3^5$  & orbital basis \\ \hline
        $E^\prime$ &    1 &   \om{}{} &  \om{}{*} &       \sign{}{1} &   \om{}{} &  \om{}{*}  & $p_{+1},d_{-2},f_{+1}$\\
                   &    1 & \om{}{*} &    \om{}{} &       \sign{}{1} & \om{}{*} &   \om{}{}  & $p_{-1},d_{+2},f_{-1}$\\[4pt]
$E^{\prime\prime}$ &    1 &   \om{}{} &  \om{}{*} &       \sign{-}{1} &   \om{-}{} & \om{-}{*}  & $d_{+1},f_{-2}$\\
                   &    1 & \om{}{*} &    \om{}{} &       \sign{-}{1} & \om{-}{*} &   \om{-}{}  & $d_{-1},f_{+2}$\\
    \end{tabular*}
    \caption{Character table of the two-dimensional representations of point group $D_{3h}$ and the corresponding complex-conjugate paired one-dimensional representation of its subgroup $C_{3h}$~\cite{aroyo2006bilbao2}. $\omega$ is given by $\operatorname{exp}(i{2\pi}/{3})$.}
    \label{tab:character_table}
\end{table*}

The similarity of the aforementioned basis sets at the valley momenta can be understood by means of a simple interference argument. Let us consider a Bloch wave function at a given momentum $\mathbf{k}$
\begin{align}
    \Psi_\mathbf{k}(\mathbf{r})=\langle \mathbf{r}|\Psi_\mathbf{k}\rangle= \sum_{\mathbf{R}} e^{i\mathbf{k}\cdot \mathbf{R}} \langle \mathbf{r}|m_{\mathbf{R}}\rangle,
    \label{eq:Bloch_wave}
\end{align}
where $|m_{\mathbf{R}}\rangle$ is an orbital with magnetic quantum number $m$ located in the unit cell corresponding to Bravais point $\mathbf{R}$. 
To prove the equivalence of different choices of the basis set for Eq.~\ref{eq:Bloch_wave} at the valley momenta, we must show that the corresponding Bloch wave functions have identical characters under the symmetry classes of the corresponding little group. 
Focusing on the $C_3$ rotation, the character $\Xi$ of the Bloch wave function of Eq.~\ref{eq:Bloch_wave} arising from an orbital $m$ at Wyckoff position $\textbf{x}$ is given by
\begin{align}
    \Xi(C_3,m,\textbf{x},\textbf{k}) =
    \varphi(C_3,\textbf{x},\textbf{k})\chi_m(C_3).
\end{align}
$\varphi(C_3,\textbf{x},\textbf{k})$ and $\chi_m(C_3)$ denote, respectively, the characters of the Bloch phase and of the orbital $m$ under a $C_3$ rotation. As the triangular Wyckoff position $1a$ is located at the origin (black filled dot), its Bloch phase remains invariant. Therefore, the character $\Xi$ of the full Bloch wave function is given only by the orbital part $\chi_m(C_3)=e^{i\frac{2\pi}{3}m}=\omega^m$. 
For the honeycomb positions, instead, a $C_3$ rotation around 1$a$ translates A and B into the corresponding ones in neighboring unit cells, as illustrated in Fig.~\ref{fig:model_new}(a).
This results in a Bloch phase difference at A and B of 
\begin{align}
    \varphi\left(C_3,A,\textbf{k}=\left\{\text{K/K}^\prime\right\}\right)&=\omega^{\mp2}=\omega^{\pm1},\\
    \varphi\left(C_3,B,\textbf{k}=\left\{\text{K/K}^\prime\right\}\right)&=\omega^{\mp1}.
\end{align}
This phase difference is reflected in the opposite sequences of colors under the effect of $C_3$ rotations depending on the valley momenta, as shown in Fig.~\ref{fig:model_new}(a).

By taking into account this additional ``Bloch angular momentum'' $\Tilde{m}_{\textbf{x,k}}=\pm1$, we obtain the condition under which both Bloch wave functions have the same $C_3$ character:
\begin{align}
    0 = (m_{t}-m_{hc}-\Tilde{m}_{\textbf{x,k}})\mod{3}.
\end{align}
This implies that any Bloch wave function of a chiral orbital $m_t\mod{3}\neq0$ on the triangular lattice transforms identical as one of $m_{hc}\mod{3}=0$ orbitals located at the A/B honeycomb sites.

Having established the equivalence on the level of the Bloch wave function symmetry, we turn now to the honeycomb connectivity in the triangular lattice. This results from a wave function interference at the A/B sites, as shown in Fig.~\ref{fig:model_new}(b): the chiral flavors $m_t$ on the triangular sites (filled black circles) contribute to the Bloch wave function with their lattice and orbital phases. Evaluating the corresponding valley state from Eq.~\ref{eq:Bloch_wave} at points A and B in real space, one gets 
\begin{align}
    \left\langle \mathbf{r}=\{\mathrm{A},\mathrm{B}\}|\Psi_{\mathbf{k}=\left\{\text{K,K}^\prime\right\}}\right\rangle&\propto \sum_{\mathbf{R}} e^{i\mathbf{k}\cdot \mathbf{R}}\langle \mathbf{r} |
    m_\mathbf{R} 
    \rangle \label{eq:all_shell}\\
    &\propto \sum_{n=0}^2 
    \big[ \omega^{\Tilde{m}_\mathbf{r,k}} \cdot \omega^{m_t} \big]^{n}  
    \label{eq:one_shell} \\
    &=3\delta_{(\Tilde{m}_\mathbf{r,k} +m_t)\text{mod} 3,0},
\end{align}
where $\Tilde{m}_\mathbf{r,k}$ reflects the winding of the Bloch phase around the considered honeycomb site at a given valley momentum.
In Eq.~\ref{eq:one_shell} we have restricted ourselves to the shell of first nearest-neighbors but the argument remains valid if farther shells are included.
Indeed, these come in groups of triangular triplets and exploits their $C_3$ symmetric arrangement, as shown in Fig.~\ref{fig:model_new}(b) for the first nearest-neighbors.

As a result, what dictates the constructive or destructive interference for a Bloch wave function on the triangular lattice at the A/B sites is the total angular momentum $M=\Tilde{m}_\mathbf{r,k}+ m_t$. Thus, the charge density is finite when total invariance under $C_3$ symmetry can be achieved i.e., at the A or B position where the total $M$ either vanishes or is a multiple of 3:
\begin{align}
    \left|\Psi_{\mathbf{k}=\{\text{K,K}^\prime\}}\left(\mathbf{r}=\{\mathrm{A},\mathrm{B}\}\right)\right|^2\begin{cases}
    >0, & \text{if } M\!\!\!\!\mod 3 = 0 \\
    =0, & \text{if } M\!\!\!\!\mod 3 \neq 0.
    \end{cases}
    \label{eq:Bloch_loc}
\end{align}
At the valley momenta, a pair of chiral orbitals $\pm m$ at the 1$a$ triangular sites can hence contribute to the electronic charge density at the 2$b$ positions A/B. Depending on the combination of local angular momentum and lattice phase, the interference in A or B can indeed be constructive, as illustrated in Fig.~\ref{fig:model_new}(b). This results in an electronic honeycomb connectivity on the 2$b$ Wyckoff positions.  Specifically, for a pair of chiral orbitals $p_\pm=(p_x\pm i p_y)/\sqrt{2}$ with $m=\pm1$ localized at the 1$a$ site, Eq.~\ref{eq:Bloch_loc} gives that $p_+(p_-)$ localizes at A(B) and B(A) for the valley momenta K and K$^\prime$, respectively.

\section{Valley Hamiltonian}
\label{sec:valley_sym_anal}
In Sec.~\ref{sec:emergent_hc}, we have established the conditions leading to a charge profile spatially displaced w.r.t. the atomic positions.
Here, we derive the time-reversal symmetric Hamiltonian for a chiral doublet at the valley momenta, where this interference phenomenon is active. Based on a group-theory analysis, we describe the consequences of spin-orbit coupling and in-plane inversion symmetry breaking (ISB), which corresponds to considering the A and B positions in Fig.~\ref{fig:model_new} inequivalent. Finally, we establish a formal mapping at K and K$^\prime$ between the triangular chiral wave functions and the basis describing graphene-like systems within the Kane-Mele model. 

In the freestanding triangular lattice, the little group of the K/K$^\prime$ points is given by $D_{3h}$~\cite{aroyo2014brillouin}. 
This comprises one threefold vertical rotation axis, three twofold horizontal rotation axes, three vertical reflection planes, and the horizontal reflection plane. As shown in the character Tab.~\ref{tab:character_table}, its two two-dimensional representations promote symmetry-protected twofold degenerate states, and these give rise to Dirac points. 

Upon introducing SOC, the little group of K/K$^\prime$ is the double group $D_{3h}^D$ of $D_{3h}$, which contains only two-dimensional spinor representations. Assuming local atomic SOC $\vec{\hat{L}}\cdot\vec{\hat{S}}$ acting on any basis function pair of $D_{3h}$ (Tab.~\ref{tab:character_table}), the valley Hamiltonian is given by
\begin{equation}
    \hat{H}^{\mathrm{SOC}}=\lambda_\mathrm{SOC}\vec{\hat{L}}\cdot\vec{\hat{S}}=\lambda_\mathrm{SOC}\hat{L}_z \otimes \hat{S}_z,
    \label{eq:H_soc}
\end{equation}
with $\vec{\hat{L}}$ and $\vec{\hat{S}}$ denoting the orbital and spin angular momentum operator, respectively.
The twofold degenerate valence and conduction eigenstates of Eq.~\ref{eq:H_soc} are
\begin{align}
    E_{v}:\hspace{1cm} |\Psi_v\rangle=\,&\left|j_z = \mp |m|\pm \frac{1}{2}\right\rangle, \\
    E_{c}:\hspace{1cm} |\Psi_c\rangle=\,&\left|j_z = \pm |m|\pm \frac{1}{2}\right\rangle.
\end{align}
As positive and negative $m$-quantum numbers contribute to both valence and conduction eigenstates, the pairs of degenerate eigenvalues localize on both A/B sublattices resulting in a honeycomb-type charge pattern.

Of central importance to our analysis, is the symmetry reduction induced by a breaking of the in-plane inversion (A and B inequivalent).
The absence of the three vertical reflection planes results in the little group $C_{3h}$~\cite{kochan2017model} and the above-mentioned twofold representations split up into pairs of chiral one-dimensional representations (see $C_{3h}$ in Tab.~\ref{tab:character_table}). 
Hence, in each spin sector, the low-energy Hamiltonian of ISB-split Dirac states can be parametrized 
by the $z$-component of the orbital angular momentum operator $\hat{L}_z$: 
\begin{align}
    \hat{H}^{\mathrm{ISB}}\left(\mathrm{K}/\mathrm{K}^\prime\right)=\pm\lambda_{\mathrm{ISB}}\hat{L}_z.
    \label{eq:isb_effective}
\end{align}
The $\pm$ sign in Eq.~\ref{eq:isb_effective} is a consequence of K and K$^\prime$ being time-reversal partners and $\lambda_\mathrm{ISB}$ denotes the strength of the ISB.
Eqs.~\ref{eq:Bloch_loc} and \ref{eq:isb_effective} imply that for a given band the charge localization on A and B is valley independent. 

The combined action of ISB and SOC determines the valley eigenspectrum
\begin{align}
    \hat{H}^\text{triang}\left(\text{K/K}^\prime\right)
    =&\hat{H}^\mathrm{SOC}+\hat{H}^\mathrm{ISB}\left(\text{K/K}^\prime\right) \nonumber\\
    =& \hat{L}_z \otimes \left(\lambda_\mathrm{SOC}\hat{S}_z \pm\lambda_\mathrm{ISB} \hat{S}_0\right),
    \label{eq:valley_ham_soc_isb}
\end{align}
where $S_0$ is the 2$\times$2 identity matrix.
The relative strength of $\lambda_\mathrm{ISB}$ and $\lambda_\mathrm{SOC}$ dictates the gap and, in turn also the pattern of charge localization. In the large $\lambda_\text{ISB}$ limit, the charge density from the two valence eigenstates localizes on only one of the two void positions (A or B) of the triangular lattice (as stated above, independently on the valley). In contrast, in the $\lambda_\text{SOC}$-dominated case, those two eigenstates contribute to the charge localization at both void positions A and B. Moreover, according to Eq.~\ref{eq:Bloch_loc}, each eigenstate contributes just to one void (either A or B).

This competition is in close analogy to the Kane-Mele model for graphene upon replacing the concept of voids with the sublattice degree of freedom. There, a large Semenoff mass favors a staggered charge density profile driving the system towards a topologically trivial phase. Not being adiabatically connected to this atomic limit, the non-trivial ground state necessarily inherits instead contributions from both sublattices~\cite{kane2005z,semenoff1984condensed}.

To put this equivalence on formal grounds, we derive in Appendix \ref{ap_sec:basis_trafo} a valley-dependent unitary transformation mapping the chiral basis on the triangular lattice used in Eq.~\ref{eq:valley_ham_soc_isb} (e.g.~$\{p_+,p_-\}$) onto the sublattice subspace of the honeycomb $\{\text{A,B}\}$. This reads
\begin{align}
    \hat{U}_{\mathrm{K}} = 
    \begin{pmatrix}
    1 & 0 \\
    0 & 1 
    \end{pmatrix}\otimes \hat{S}_0, \quad
    \hat{U}_{\mathrm{K}^\prime} = 
    \begin{pmatrix}
    0 & 1 \\
    1 & 0 
    \end{pmatrix}\otimes \hat{S}_0,
    \label{eq:valley_transform}
\end{align}
and transforms the Hamiltonian in Eq.~\ref{eq:valley_ham_soc_isb} onto
\begin{align}
    \hat{H}^\mathrm{KM}\left(\mathrm{K/K}^\prime\right)
    &=\hat{U}_{\mathrm{K/K}^\prime}^\dagger \hat{H}^\text{triang}\left(\text{K/K}^\prime\right) \hat{U}_{\mathrm{K/K}^\prime} \nonumber\\
    &=\pm \lambda_\mathrm{SOC}\tau_z\otimes \hat{S}_z + \lambda_\mathrm{ISB}\tau_z \otimes \hat{S}_0,
    \label{eq:H_valley_rot}
\end{align}
where $\vec{\tau}$ denote Pauli matrices representing the sublattice degree of freedom. To give specific examples, a $\{p_+,p_-\}$ basis can be mapped onto a $\{s_\text{A},s_\text{B}\}$-like honeycomb basis located on the sublattices A and B. A $p_z$-like graphene basis would be instead obtained starting from a triangular $d_\pm=(d_{xz}\pm i d_{yz})/\sqrt{2}$ basis (odd under reflections w.r.t. the horizontal plane, see Appendix \ref{ap_sec:basis_trafo}). 

The local atomic SOC term of the triangular basis is hence transformed into a non-local Kane-Mele-type SOC interaction, while the valley-dependent triangular ISB term turns into a local staggered potential/Semenoff mass~\cite{kane2005z,semenoff1984condensed}. It is interesting to note, that the strength of the Kane-Mele SOC interaction in Eq.~\ref{eq:H_valley_rot}, which in graphene-type systems originates from intrinsically weak second-neighbor processes, is here as large as the local atomic SOC~\cite{bauernfeind2021design,reis2017bismuthene}.


We have therefore proven the existence of a transformation between the honeycomb and the triangular basis sets. For the latter, we will now show that not only does the interference mechanism displace the charge centers away from the atomic positions when focusing on the valley momenta, but we can even define localized obstructed Wannier functions for the whole BZ.
While the existence of such Wannier functions does not come as a surprise when the ISB term is large and the system is a trivial insulator, the possibility of constructing them in the SOC-dominated non-trivial limit is by no means obvious.

\section{Obstructed QSH-insulator}
\label{sec:tb_section}
\begin{figure}[b]
    \centering
    \includegraphics[width=1.0\linewidth]{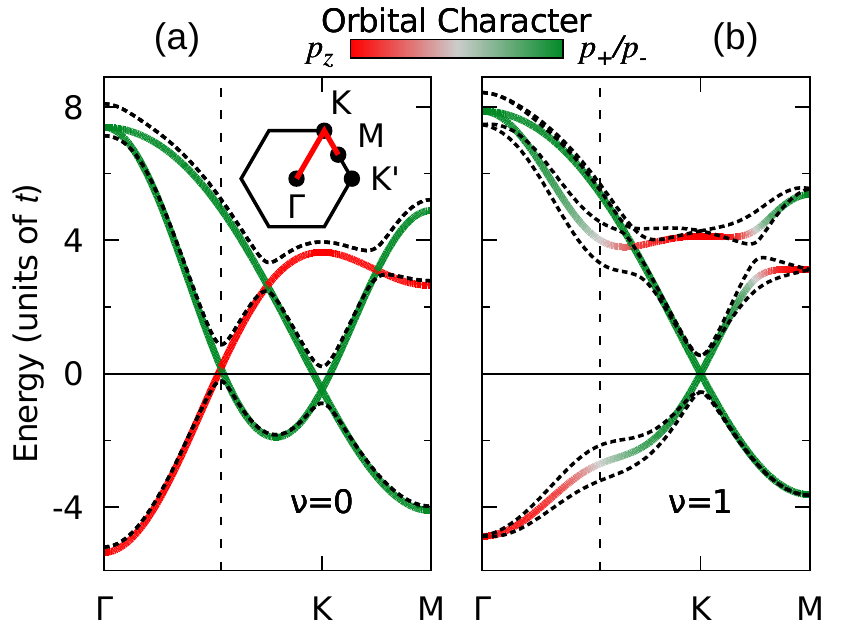}
    \caption{Band structure of the $\{p_\pm, p_z \}$ model on the freestanding (a) triangular lattice and in the absence of a mirror symmetry with respect to the horizontal plane (b). The color code of the solid lines denotes the orbital character for the model without SOC while the bands with SOC are plotted with thick dashed lines. The thin dashed vertical lines indicate the position of the $p_\pm$-$p_z$ degeneracy (in the absence of SOC and mirror symmetry breaking) and the inset to (a) shows the path in the BZ. The insulating phase with SOC for the freestanding case, see thick dashed line in (a), is topologically trivial ($\nu$=0) whereas, the concomitant presence of SOC and of the mirror-symmetry breaking, thick dashed lines in (b), turns it is into a QSHI ($\nu$=1).}
    \label{fig:model_no_isb}
\end{figure}
We start from the simplest tight-binding description of a $p$-shell on the triangular lattice possessing only nearest-neighbor hopping. 
In addition to the Dirac point at K/K$^\prime$, the Hamiltonian for $\{p_+,p_-\}$ orbitals displays twofold degenerate eigenvalues also at the $\Gamma$ point, making the band structure inevitably metallic \footnote{The band structure can be at most semi-metallic, in the special case in which the eigenvalues at $\Gamma$ are energetically aligned with those at the valley momenta.}, as shown by the green bands in Fig.~\ref{fig:model_no_isb}(a). This is therefore different from typical graphene-like systems (two sublattices and one orbital per site only), in which the Dirac cone lives in an otherwise already gapped band structure. 
This situation is not changed qualitatively by the $p_z$ ($m$=0) orbital, shown by red solid lines in Fig.~\ref{fig:model_no_isb}), which introduces additional crossings between the in-plane and out-of plane degrees of freedom.
However, an insulating $\nu=0$ ground state can be stabilized in the presence of local SOC (see dashed bands in Fig.~\ref{fig:model_no_isb}) as this couples the two subspaces, gapping out all low-energy degeneracies.

The band crossing marked by the dashed vertical line approximately half-way between $\Gamma$ and K, which has turned into an avoided crossing after having swiched on SOC [black dashed lines in Fig.~\ref{fig:model_no_isb}(a)] is strongly sensitive to the breaking of the horizontal mirror reflection symmetry. This is apparent by looking at Fig.~\ref{fig:model_no_isb}(b).
By breaking this symmetry, hoppings between the $\{p_+,p_-\}$ and the $p_z$ orbitals are not longer symmetry-forbidden and, as a consequence, a band structure with Dirac cones living in a global gap can be obtained (see Appendix \ref{ap_sec:tb_model} for more details on the microscopic Hamiltonian). 
This hybridization is reminiscent of Rashba systems, in which the broken 
mirror symmetry allows for the overlap of the radial in-plane and the $p_z$ wave functions by promoting in-plane OAM polarization (see vertical dashed lines in Fig.~\ref{fig:model_no_isb}), the OAM texture in the full Brillouin zone (BZ) is shown in Fig.~\ref{fig:L_map}. Distinguishing the in-plane orbital for their tangential and radial alignment clarifies also why one crossing in the conduction bands remains [along $\overline{\Gamma\text{K}}$, Fig.~\ref{fig:model_no_isb}(b)] as it is due to the tangential in-plane component~\cite{petersen2000simple,unzelmann2020orbital}.

Having defined a tight-binding model in the full BZ, we focus in the following on the two aspects demonstrating 
that the {$\nu$=$1$} phase of our model can be seen as an obstructed QSHI: ($i$) the non-representability in terms of time-reversal symmetric Wannier functions and ($ii$) the displacement of the centers of a pair of Kramers-violating Wannier functions away from the lattice positions.
To this aim, we introduce trial Wannier basis sets $\ket{\tau_i}$ and calculate the overlap matrix with the occupied Bloch bands~\cite{Soluyanov2011wannier}:
\begin{align}
    S_{ij}(\mathbf{k})=\braket{\tau_i|\hat{\mathcal{P}}(\mathbf{k})|\tau_j},
\end{align}
where $\hat{\mathcal{P}}$ is the projector onto the occupied states
\begin{align}
    \hat{\mathcal{P}}(\mathbf{k}) =  \sum_n^N\ket{\Psi_{n\mathbf{k}}}\bra{\Psi_{n\mathbf{k}}}.
\end{align}
Hence, the absence of vanishing overlap eigenvalues of $S(\mathbf{k})$ in the full BZ indicates the representability of the valence bands in terms of the Wannier trial basis.

\begin{figure}
    \centering
    \includegraphics[width=\linewidth]{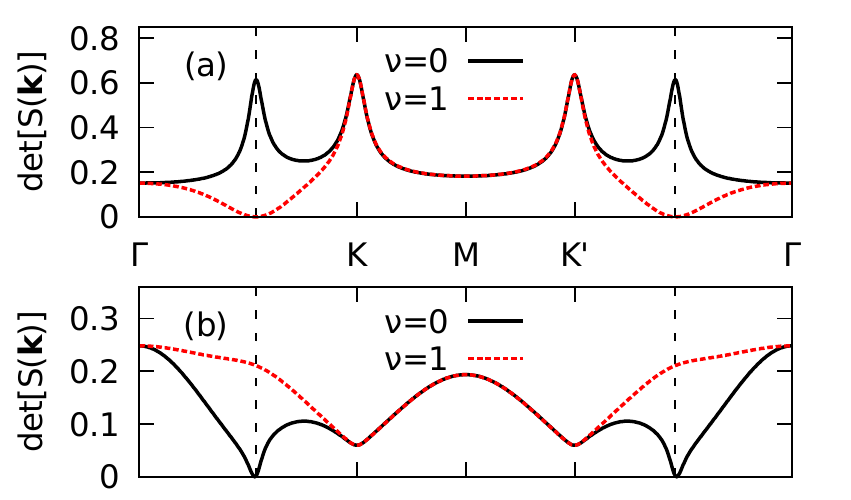}
    \caption{det$[S(\textbf{k})]$ along the diagonal of the BZ for a trial basis of $\ket{j,j_z}=\{\ket{1/2,\pm1/2}$ orbitals on the triangular site (a) and $\ket{sp_z^A,+},\ket{sp_z^B,-}$ orbitals on the interstitial sites (b). The dashed vertical lines indicate the position of the crossing of the $p_\pm$ and the $p_z$ band in the absence of SOC and mirror symmetry breaking.}
    \label{fig:det_S_path}
\end{figure}

To illustrate point ($i$), we construct trial functions for the valence band of the $\nu$=$0$ phase, in which all low-energy metallic band crossings are gapped by the local SOC [see Fig.~\ref{fig:model_no_isb}(a)]. First, we choose a time-reversal symmetric pair of total angular momentum eigenfunctions $\ket{j,j_z}=\ket{1/2,\pm1/2}$. As shown by the black solid curve in Fig.~\ref{fig:det_S_path}(a), det$[S(\textbf{k})]$ is finite everywhere, indicating the absence of vanishing overlap eigenvalues. For a sufficiently strong in-plane/out-of plane hybridization, i.e. in the presence of a vertical reflection symmetry-breaking, the model features a large gap at the momenta marked by the vertical dashed lines in Fig.~\ref{fig:model_no_isb}(b), and is stabilized in a non-trivial $\nu$=$1$ phase. We have confirmed this also through an explicit calculation of the Wilson loop eigenvalues, following Refs.~\cite{soluyanov2011computing,yu2011equivalent}. 
In this topological phase, the ground state cannot be represented using Wannier Kramers' pairs.
The overlap indeed vanishes at points in the BZ [see dashed red line in Fig.~\ref{fig:det_S_path}(a)]. This happens right at the above-mentioned momenta between $\Gamma$ and K at which the gap is opened by mirror symmetry-breaking (dashed vertical lines). 
As shown by Soluyanov and Vanderbilt in Ref.~\cite{Soluyanov2011wannier} for the Kane-Mele model, a localization can instead be achieved upon introducing two time-reversal symmetry-violating Wannier functions
\begin{align}
    \ket{\tau_i}=\{\ket{A,\uparrow_x},\ket{B,\downarrow_x}\},
    \label{eq:Sol_AB_proj}
\end{align}
which are localized on the two honeycomb Wyckoff positions with an in-plane spin-alignment in the $x$ direction.

A representability of the valence bands of our triangular model in this non-atom centered Wannier basis would indicate a real-space obstruction. To account for the presence of even and odd band character under reflections w.r.t. an horizontal plane (see Fig.~\ref{fig:model_no_isb}) we choose a $sp_z$ hybrid orbital on the A and B sites
\begin{align}
    \ket{\text{A/B}} = \frac{1}{\sqrt{2}}\left(\ket{s}+\ket{p_z}\right).
    \label{eq:spz_hyb_proj}
\end{align}
As shown in Fig~\ref{fig:det_S_path}(b), this trial basis results in finite overlap eigenvalues for the $\nu$=$1$ phase and overlaps for the $\nu$=$0$ phase which instead vanish at points in the BZ (see also Fig~\ref{fig:det_S_map}). This is at odds with the case of symmetry-preserving trial wave functions, as shown in Fig~\ref{fig:det_S_path}(a). Again, these decisive differences in det$[S(\textbf{k})]$ can be seen at the momenta of the lifted $p_\pm$-$p_z$ degeneracy, indicating its importance for the real-space obstruction: If mirror symmetry breaking dominates over SOC, the valence bands share the same in-plane OAM polarization and their local orbital angular momentum vector covers the whole unit sphere in the full BZ (see also Sec.~\ref{ap_sec:obstruction} and Fig.~\ref{fig:L_map}). Hence, a trial basis on the triangular positions with non-vanishing overlap eigenvalues cannot exist for the $\nu$=1 phase.


Further it should be noted, that despite the different $\mathbb{Z}_2$-invariant, both phases have identical irreducible band representations at the high symmetry points. The gap reopening occurs indeed at the non-high symmetry momenta marked by the vertical dashed lines. For this reason, the topological invariant cannot be obtained from symmetry indicators~\cite{bradlyn2017topological}. 

\section{Experimental confirmation of orbital character}
\begin{figure}[t]
    \centering
    \includegraphics[width=1.0\columnwidth]{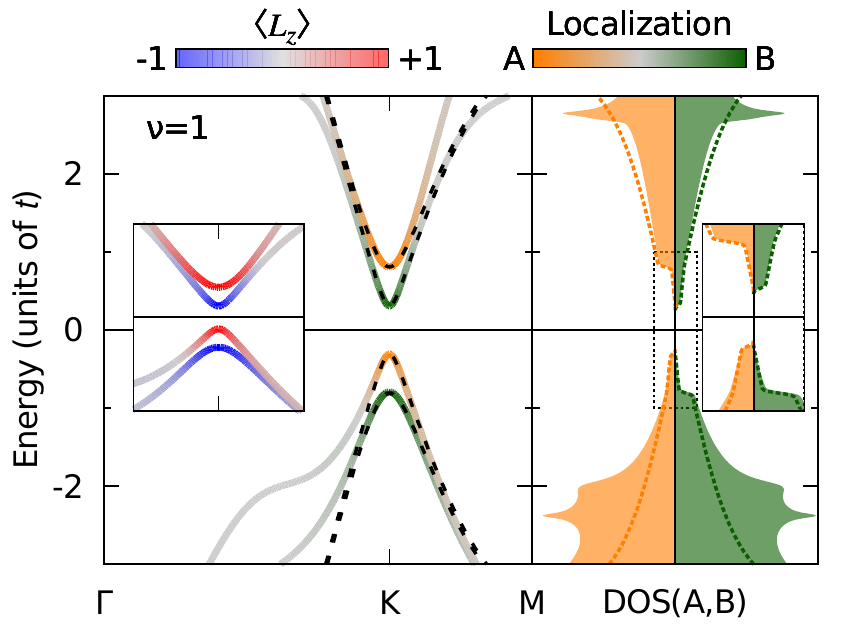}
    \caption{Band structure and sublattice projected DOS of the triangular and Kane-Mele model in the non-trivial phase.
    The solid lines/filled curve correspond to the triangular lattice, whereas the dashed lines to the Kane-Mele model. The blue-red color code denotes the OAM character (inset to the band structure), while the green-orange one encodes the A/B localization. The dashed box indicates the region shown in the inset to the sublattice projected DOS.}
    \label{fig:model_isb_KM}
\end{figure}
In the following, we will elaborate on the consequences of the obstruction of our model in terms of wave function symmetry and real-space localization. This will allow us to prove the existence of obstructed QSHIs based on experiments on a recently discovered triangular Dirac system~\cite{bauernfeind2021design,erhardt2022indium}.

\begin{figure}[b]
    \centering
    \includegraphics[width=1.0\columnwidth]{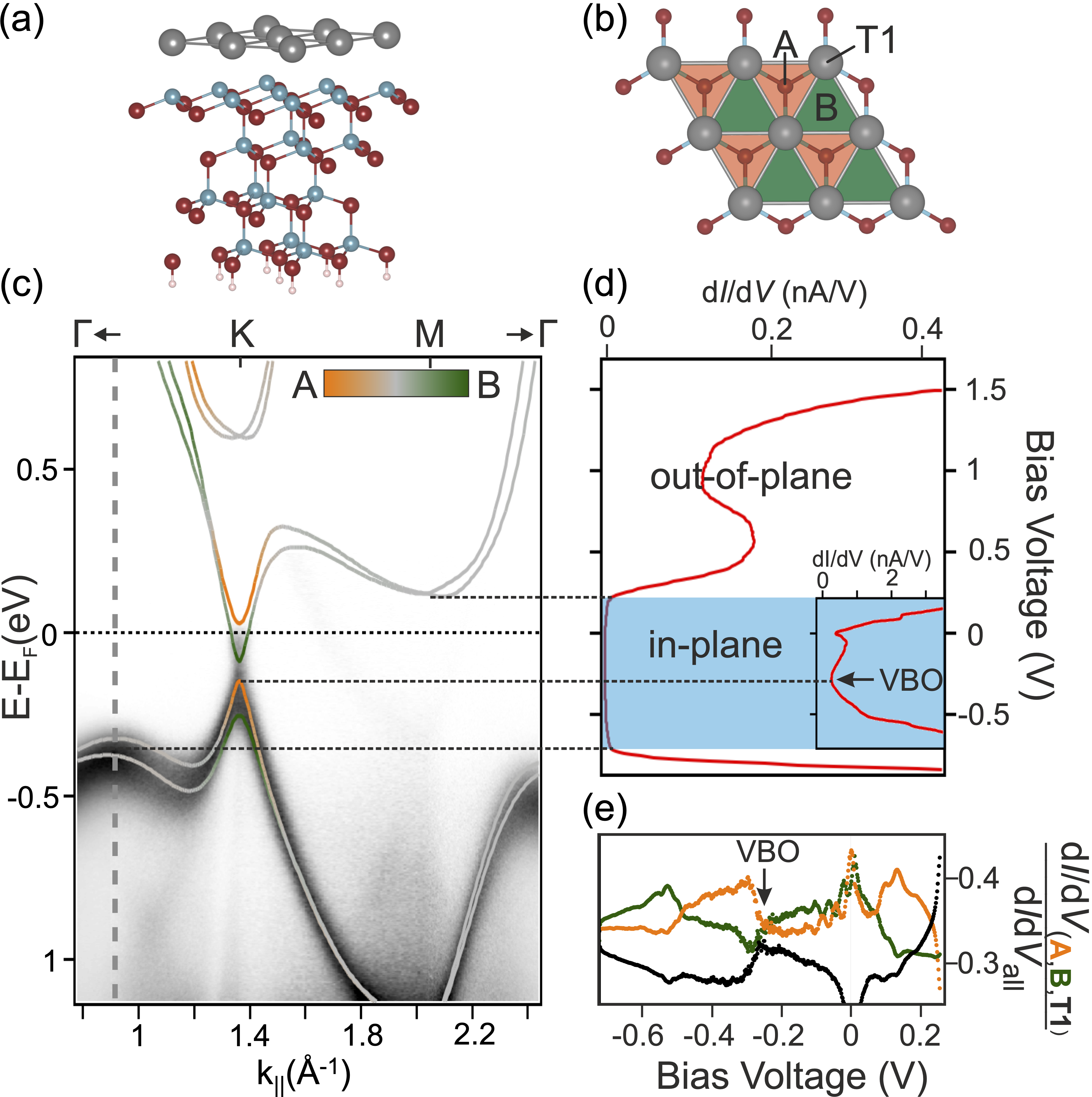}
    \caption{Side (a) and top view (b) on the unit cell: Gray and cyan spheres represent indium and silicon atoms, respectively. Carbon atoms are denoted in red and the pink spheres correspond to the hydrogen atoms used for passivation~\cite{vesta}; just for the clarity of the illustration, we show here the topmost SiC layer only. (c) Comparison of ARPES measurements of indenene and $G_0W_0$ band structure along the  $\Gamma$-$\mathrm{K}$-$\mathrm{M}$-$\Gamma$ path. (d) STS measurements stabilized at tip-sample distance z$_0$ and z$_1$=z$_0$-7.8$\, \text{\AA}$ (inset). At z$_0$ STS is less sensitive to in-plane states effectively probing only $p_z$-like indenene states. High $n$-type doping of the substrate shifts the valence band onset (VBO) to approximately -300$\,$mV aligning with our ARPES data. (e) Renormalized $\dd I/\dd V$ signal recorded at the A (orange), B (green) and T1 (black) site of the indenene unit cell.  
    }
    \label{fig:GW_exp}
\end{figure}

Without loss of generality, we consider hereinafter a small ISB term (see Eq.~\ref{eq:valley_ham_soc_isb}), which lifts the degeneracy of the SOC gapped Dirac bands by promoting out-of-plane orbital angular momentum as shown for the {$\nu$=$1$} phase in Fig.~\ref{fig:model_isb_KM}. As a consequence of SOC, orbital and spin degrees of freedom are mixed, therefore the valence doublet is formed by a $p_+$ and a $p_-$ state (see inset). This localizes at opposite A/B positions (see Eq.~\ref{eq:Bloch_loc}) resulting in an honeycomb-like electron charge-density profile displaced from the 1$a$ Wyckoff- to the A/B positions. This is further confirmed by the sublattice-projected DOS in the right panel of Fig.~\ref{fig:model_isb_KM}.
A similar band structure has been realized in indenene~\cite{bauernfeind2021design}, a $p$-electron triangular monolayer of indium atoms grown on a SiC substrate [Fig.~\ref{fig:GW_exp}(a,b)]. This represents therefore the ideal system to experimentally confirm our theoretical analysis.
Angle-resolved photoelectron spectroscopy (ARPES) and {\it ab-initio} electronic structure of indenene, within the framework of GW many-body perturbation theory, are shown in Fig.~\ref{fig:GW_exp}(c) (for details see appendix Secs~\ref{ap_sec:GW}, \ref{ap_sec:ARPES}, \ref{ap_sec:STM_STS} and Ref.~\onlinecite{bauernfeind2021design}).
One immediately recognizes the presence of all fundamental ingredients we have introduced in Sections
\ref{sec:emergent_hc} to \ref{sec:tb_section}: a sizable hybridization gap opened between the radial in-plane and the $p_z$-band and four eigenvalues at the valley. 
The energy dispersion of the Dirac fermions is attributed to the interplay of the relatively strong local SOC of the indium atoms ($\lambda_{\mathrm{SOC}} = 0.43$ eV) and the ISB is induced by the specific (T\textsubscript{4}) arrangement of the carbon atoms in the surface layer of SiC [see Fig.~\ref{fig:GW_exp}(b)].
Further, the presence of the SiC substrate opens a $p_z$-$p_{\pm}$ hybridization gap of approximately 1$\,$eV leaving only the Dirac bands around the Fermi level.
We address the orbital-character of these bands by performing orbital-symmetry selective scanning tunneling spectroscopy (STS), i.e. by varying the tip-to-sample distance~\cite{FeenstradIdVk}.  
A measurement at large distances -- main panel of Fig.~\ref{fig:GW_exp}(d) -- probes mainly the out-of-plane 
contribution of the indium states. 
Here, striking signature of the absence of low-energy $p_z$-states is drawn from the sudden drop of the d$I$/d$V$ signal aligning with the $p_z$-$p_{\pm}$-hybridization maxima of our GW calculations (horizontal dashed lines in panel c).
Small tip-to-sample distances instead reveal also a finite local DOS (LDOS) inside the $p_z$-$p_{\pm}$-hybridization gap, a clear signature of the in-plane Dirac feature contributing to these energies (inset to panel d)~\cite{bauernfeind2021design}.
The corresponding lattice site-resolved (A, B and T1) LDOS-mapping [see Fig.~\ref{fig:GW_exp}(e)] shows dominant differential conductance at T1 for $>$0.23$\,$V where the steep out-of-plane onset appears in panel d. This is excellent agreement with our theoretical modelling predicting a strong $p_z$-component at the same energy (see again horizontal dashed line between panel c and d). 
On the contrary, probing only the Dirac states between -0.72$\,$V and 0.23$\,$V, the charge maximum localizes at either site A or B, displaced away from the lattice position. Given our theoretical understanding of the interference mechanism, this displacement hints at the existence of a real-space obstruction.
In particular, the charge maximum alternates between A and B following the {$\nu$=$1$} energy sequence of Fig.~\ref{fig:model_isb_KM} and thus puts forward indenene as an obstructed QSHI.

\section{Edge states of obstructed QSH-insulators}

\begin{figure}[t]
    \centering
    \includegraphics[width=\columnwidth]{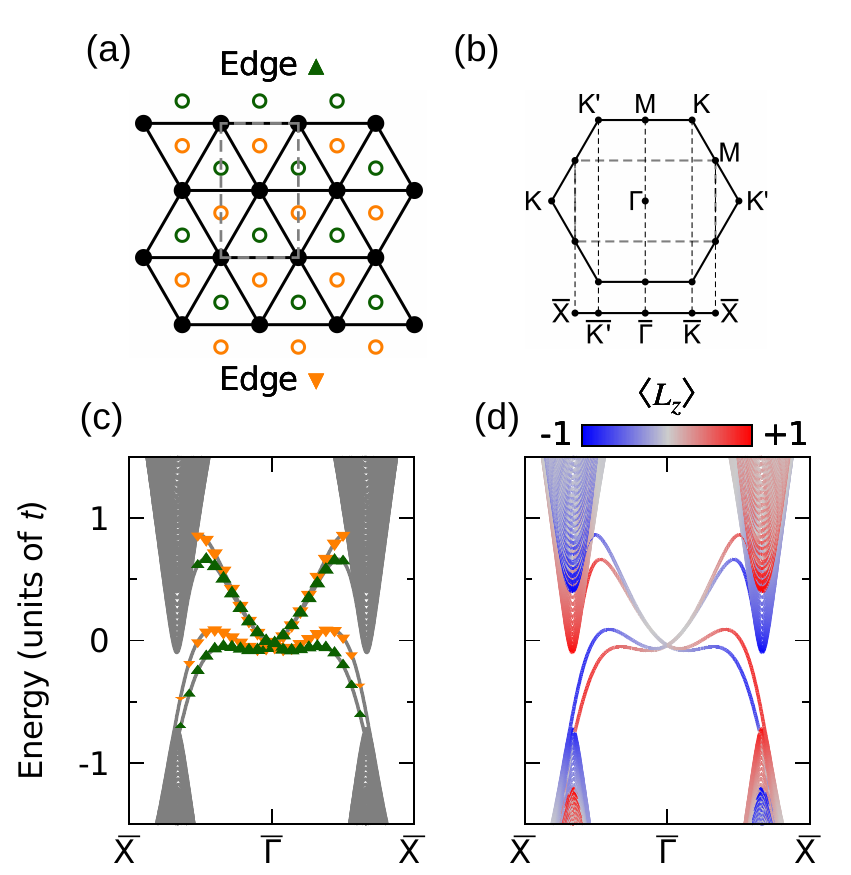}
    \caption{
    (a) conventional unit cell (gray) and edge geometry.
    (b) 2D BZ and backfolding of high symmetry momenta to 1D slab BZ. Edge localization (c) and OAM polarization (d) in the slab model.}
    \label{fig:slab_fig}
\end{figure}

It has been shown that in trivial insulators the real-space obstruction can be accompanied by the presence of in-gap surface modes~\cite{obstructed1,obstructed2,obstructed3}. Hence, it is relevant to inspect the edge modes of the obstructed QSHI, which are guaranteed to exist by the bulk-boundary correspondence. We will then compare them to those of its non-obstructed analogue, represented in the present case by the Kane-Mele model.

However, before addressing a slab geometry, we first compare the two models at the level of the bulk properties. This can be done by inspecting the band structure of the two models, which we show in the left panel of Fig.~\ref{fig:model_isb_KM} for the non-trivial phase. 
Close to the valley momenta, the dispersion of the obstructed and non-obstructed QSHI show no qualitative difference.
This is true as long as non-linear contributions for the multi-orbital triangular lattice are negligible. At higher energies, the interplay of SOC and in-plane and out-of-plane polarization flattens the bands. Though not directly connected with the real-space obstructed nature, it is interesting to note that this promotes van-Hove-singularities, which at these energies are absent in the single-orbital Kane-Mele model (see right panel of Fig.~\ref{fig:model_isb_KM}).

We now turn to the helical edge modes of the triangular QSHI and highlight how they signal the existence of the real-space obstruction.
This manifests itself in a striking way in ribbon geometries, considering the termination in which A and B are located on opposite edges, see Fig.~\ref{fig:slab_fig}(a).
Here, we therefore focus on the ``flat'' edge on the top and bottom sides of the sketch and do not further discuss the other ``zigzag''-like edge, shown on the left and right sides.
The color and the point size in Fig.~\ref{fig:slab_fig}(c) indicates the spatial localization of the states on the top and bottom edges. The fact that the edge states from opposite terminations are not degenerate indicates how they inherit also a strong A/B localization character, which is different due to the presence of ISB. This character is particularly pronounced close to the valley momenta (see also Fig.~\ref{fig:model_isb_KM}), where we know from Eq.~\ref{eq:Bloch_loc} that the $p_\pm$ Bloch states localize at the A/B voids.
Hence, we identify an observable signature of the obstructed nature of this QSHI.

Further, differently from the Kane-Mele model where the pure $p_z$ ($m$=0) character forbids a finite OAM, the obstructed QSHI on the triangular lattice features edge states that are OAM polarized, see Fig.~\ref{fig:slab_fig}(d).
By moving away from the valley momenta towards $\Gamma$, the $p_z$ degree of freedom in the edge states becomes more and more dominant. As a result of the $\vec{L}\cdot\vec{S}$ coupling, $\langle L_y\rangle$ as well as $\langle S_y\rangle$ become finite, rotating the OAM and the spin into the $yz$-plane.
Finally, due to the strong non-linearity away from the valley momenta, the dispersion of the edge states as well as their crossing at $\overline\Gamma$ deviate strongly from the case of the chiral-symmetric Kane-Mele model.



\section*{Conclusion}

In this work, we have established that the triangular and the honeycomb lattice can be seen as coequal partners on the level of their Dirac fermions. Extending this concept to the full BZ, we find that the triangular lattice can host graphene-like valence bands arising from time-reversal symmetry-breaking Wannier functions on the honeycomb Wyckoff positions, i.e. at the voids of the triangular lattice. This generalizes the notion of real-space obstruction recently developed for trivial insulators and hence also its physical consequences to quantum spin-Hall systems. Further, it allows for a connection with the higher-order topological classification, distinguishing between systems with the same non-trivial $\mathbb{Z}_2$-invariant.

One decisive difference that emerges from our analysis, is that a QSHI phase on the triangular lattice requires, in contrast to Kane-Mele-type systems, a sufficiently strong potential which breaks the horizontal mirror reflection.
In combination with the full-atomic SOC, this promotes a local OAM winding on the triangular site, which prohibits the existence of a non-obstructed Wannier representation. Our experiments on indenene, the only triangular QSHI realized so far, support our theoretical prediction and make this topological monolayer a potential candidate for future experiments on obstructed QSHIs.

Since QSHIs lack a Wannier representation, the widely used high-throughput-suited symmetry indicator schemes cannot be applied for the detection of real-space obstructed non-trivial phases.
Upon releasing the constraint of internal symmetry-preserving Wannier functions, Soluyanov-Vanderbilt-like representations can be constructed to determine the real-space localization.
Hence we expect that a sizable number of already known non-trivial $\mathbb{Z}_2$-indicated systems may belong to this new class of topological materials. As witnessed by our triangular model, the false-negatives of TQC appear to be a potential pool of candidates.

Concerning future investigations, it must be noted that the real-space obstruction of a QSHI can be probed at lattice defects of heterostructures in which the obstructed QSHI in question is completely surrounded by a non-obstructed one. The helical edge states of the two QSHIs gap each other out. In the L-shaped corner regions~\cite{Kooi2021bulkcorner} a quantized corner charge (possibly accompanied by in-gap corner modes) is then expected to appear.
Obstructed QSHIs can be also used as basic building blocks of three-dimensional topological crystalline phases. Consider for instance a three-dimensional bulk crystal with a ${\mathcal C}_{n z}$ rotation symmetry where the order of the rotational symmetry $n=2,4,6$. The three-dimensional BZ of this system can be viewed as a collection of two-dimensional cuts parametrized by the momentum $k_z$ parallel to the rotation axis. Assuming that QSHIs are realized at the time-reversal invariant cuts $k_z=0,\pi$, the bulk crystal must represent a weak three-dimensional topological insulator. Let us further consider that out of the two $k_z=0,\pi$ cuts, one is an obstructed QSHI. When subject to translational symmetry breaking perturbations that double the stacking period, the system will be transformed in a topological crystalline insulator with anomalous surface Dirac cones connected by helical hinge modes~\cite{fang2019new}. Additionally, the system might also realize an hybrid weak topological insulator~\cite{Kooi2020hybridorder} with rotational symmetry-protected Dirac cones on its so-called dark surfaces.

Besides the intriguing consequence of real-space obstruction, the realization of a certain Bloch wave function symmetry on different crystal lattices constitutes a conceptual and practical difference in the microscopic origin of the relevant hopping processes. Specifically, the 2$^\text{nd}$-nearest-neighbor Kane-Mele SOC and the local staggered potential (Semenoff mass) in the honeycomb correspond to the local SOC of Eq.~\ref{eq:H_soc} and the nearest-neighbor ISB terms of Eq.~\ref{eq:isb_effective} in the triangular lattice, respectively. This means that the resulting graphene-like Dirac fermions are gapped by an atomic $\vec{L}\cdot \vec{S}$-type SOC interaction, which is in general much stronger than its non-local equivalent~\cite{reis2017bismuthene,bauernfeind2021design}.
As a consequence, the valley Hamiltonians in the two basis sets have an identical structure, but can happen to live in completely different parameter regimes because of different physical origins of the associated interactions.

From a general perspective, realizing a specific wave function symmetry with different basis sets in a periodic geometry will not only stimulate the search for topological materials but may also pave the way to new approaches for the investigation and realization of physical phenomena in a different context, e.g., cold atoms in optical lattices, photonic crystals, or acoustic lattices. This could help to overcome technical challenges in terms of realizability and may also give access to new parameter and phase regimes. Further, the $\mathbf{k}$-dependent basis mapping introduced here will have non-trivial implications for interacting particles. In that case, the transformation would indeed involve also two-body terms and may hence lead to the appearance of interaction terms that differ substantially from those considered in standard cases.

\bigskip
{\noindent
	\textbf{Acknowledgements}
	We acknowledge helpful discussions with J. Cano and Y. Fang.
	The authors are grateful for funding support from the Deutsche Forschungsgemeinschaft (DFG, German Research Foundation) under Germany's Excellence Strategy through the W\"urzburg-Dresden Cluster of Excellence on Complexity and Topology in Quantum Matter ct.qmat (EXC 2147, Project ID 390858490) as well as through the Collaborative Research Center SFB 1170 ToCoTronics (Project ID 258499086). The research leading to these results has received funding from the European Union's Horizon 2020 research and innovation programme under the Marie Sk\l{}odowska-Curie Grant Agreement No. 897276. We gratefully acknowledge the Gauss Centre for Supercomputing e.V. (www.gauss-centre.eu) for funding this project by providing computing time on the GCS Supercomputer SuperMUC-NG at Leibniz Supercomputing Centre (www.lrz.de).
}


\section*{Appendix}


\setcounter{section}{0}
\renewcommand\thesection{\Alph{section}}
\section{Valley-dependent basis transformation}
\label{ap_sec:basis_trafo}

\renewcommand{\arraystretch}{1.5}
\setlength{\tabcolsep}{18pt}
\begin{table*}[]
    \centering
    \begin{tabular}{c|c|c|c|c|c|c|c}
         $E_z$ & $V^\sigma$ & $V^\pi$ & $V_{p_z}^\pi$ & $\lambda_{\rm SOC}$  & $\lambda_{\rm MIR}$  & $\lambda_{\rm ISB}(\nu=0)$ & $\lambda_{\rm ISB}(\nu=1)$   \\
         \hline
          -1.5 &          2 &    -0.25 &         -1    &            1      &          1           &          0.75              &         0.25
    \end{tabular}
    \caption{Tight binding parameters in units of $t$.}
    \label{tab:tb_param}
\end{table*}

Here we demonstrate the equivalence of a pair of chiral orbitals on the triangular lattice and a bipartite basis on the honeycomb lattice by deriving a complete basis transformation at the valley momenta. Its existence can be proven by projecting the Bloch wave function onto Coulomb Sturmians~\cite{rotenberg1962application}, a full basis in the $\mathcal{R}^3$ which are given by
\begin{align}
    \chi_{\tau}(\mathbf{x})=R_{nl}(r)Y_{l}^m(\theta,\phi),
\end{align}
defined by a set of atomic-like quantum numbers $\tau=[n,l,m]$ centered around $\mathbf{r}_0$ with the distance vector $\mathbf{r}=\mathbf{x}-\mathbf{r}_0$. For the sake of simplicity, we will neglect in the following the the radial part $R_{nl}$ and consider only the spherical harmonics $Y_{l}^{m}$. First, we express the initial orbital at $\mathbf{r}_0$ in Coulomb Sturmians $|w_{R_n}\rangle=\sum_{\tau} c_\tau |\chi_\tau\rangle$. The transformed orbital $|w_{n^\prime}\rangle$ centered at site $\mathbf{r}_0^\prime$ is given by the projection of the Bloch wave function onto the Coulomb Sturmian basis $|\chi_{\tau^\prime}\rangle$ in the home unit cell.
\begin{align}
    |w_{n^\prime}\rangle&=\sum_{\tau^\prime} |\chi_{\tau^\prime}\rangle \langle \chi_{\tau^\prime}|  \Psi \rangle 
    \label{eq:ShiWul_proj}\\
    &=\sum_{\mathbf{R},\tau,\tau^\prime} c_\tau e^{i\mathbf{k}\cdot\mathbf{R}} |\chi_{\tau^\prime}\rangle \langle \chi_{\tau^\prime}| \chi_\tau (\mathbf{R})\rangle \\
    &\propto \sum_{\mathbf{R},\tau,\tau^\prime} e^{i\mathbf{k}\cdot\mathbf{R}} c_\tau  |Y_{\tau^\prime}\rangle\langle Y_{\tau^\prime} |Y_{\tau}(\mathbf{R})\rangle.
    \label{eq:ShiWul_Bloch}
\end{align}
The spherical harmonics are parametrized by $Y_{l}^m=P_l^m(\theta)e^{im\phi}$ with the Legendre polynomial $P_l^m$ and the spherical coordinates $(\theta,\phi)$. When transforming from position $1a$ to one of the the A/B sites, all neighbors of the same order come in triangular triplets $t$, the complex phase transforms as
\begin{align}
    &\frac{1}{3}\sum_{\mathbf{R}\in t,} e^{i\mathbf{k}\cdot\mathbf{R}} \langle Y_{\tau^\prime} |Y_{\tau}(\mathbf{R})\rangle\\ 
    \stackrel{\mathbf{k}=\text{K/K}^\prime}{\propto}& \frac{1}{3}\sum_{n}^3 e^{i\frac{2\pi}{3} n\left(\Tilde{m}_{\text{K/K}^\prime}-m^\prime +m\right)} \label{eq:trip_proj}\\
    =& \delta_{(m^\prime-m)\text{mod}(3),\Tilde{m}_{\text{K/K}^\prime}},
\end{align}
where the Bloch lattice phase enters at K/K$^\prime$ with $\Tilde{m}_{\text{K/K}^{\prime}}=\{\pm1,\mp1\}$ at $\{\text{A,B}\}$. Akin to the Bloch localization in Eq.~\ref{eq:Bloch_loc}, this relates $m$ and $-m$:
\begin{align*}
    \text{$A$:}&\quad
    \begin{cases}
    \mathbf{k}=\text{K}, &\text{if }(m^\prime-m)\mod{3}=-1 \\
    \mathbf{k}=\text{K}^\prime, &\text{if }(m^\prime-m)\mod{3}=+1
    \end{cases},\\
    \text{$B$:}&\quad
    \begin{cases}
    \mathbf{k}=\text{K}, &\text{if }(m^\prime-m)\mod{3}=+1 \\
    \mathbf{k}=\text{K}^\prime, &\text{if }(m^\prime-m)\mod{3}=-1
    \end{cases}.
\end{align*}
A further constraint arises from the symmetry of the Legendre polynomials requiring that $P_m^l$ and $P_{m^\prime}^{l^\prime}$ are both even or odd w.r.t. horizontal reflection
\begin{align*}
    |\langle P_{m^\prime}^{l^\prime} | P_m^l (\mathbf{R})\rangle| 
    \begin{cases}
    >0, \; & \text{if } (l-m+l^\prime-m^\prime)\,\text{mod 2} = 0 \\
    =0, \; & \text{if } (l-m+l^\prime-m^\prime)\,\text{mod 2} = 1
    \end{cases}.
\end{align*}

This shows indeed, that the valley Bloch function of a chiral triangular doublet $|\pm m\rangle$ with $ m\,\,\text{mod}\,\,3\neq0$ can be mapped onto a bipartite honeycomb basis whose magnetic quantum numbers are constrained to $m^\prime\,\,\text{mod}\,\,3=0$. For example a $\{p_+,p_-\}$ basis can be mapped onto a $\{s_\text{A},s_\text{B}\}$-like honeycomb basis located on the sublattice sites A and B. Consequently, a triangular $\{d_{+},d_{-}\}$ basis (odd under reflections at the horizontal reflection plane) transforms into a $p_z$-like graphene basis. The concrete basis transformation involves the elaborate evaluation of the overlap of Coulomb-Sturmians, so-called Shibuya-Wulfman integrals~\cite{avery2004many}.

\section{tight-binding model}
\label{ap_sec:tb_model}

We consider a $p$-shell $\{p_x,p_y,p_z\}$ on a triangular lattice spanned by the vectors $\mathbf{a}_1=(1,0)$ and $\mathbf{a}_2=(-0.5,\sqrt{3}/2)$. Their overlap integrals can be obtained by following the approach of Slater and Koster~\cite{slater1954simplified}:

\begin{align}
	\langle p_i| H | p_i\rangle &= n_i^2V^\sigma+(1- n_i^2)V^\pi, \\
	\langle p_i| H | p_j\rangle &= -n_i n_j (V^\pi-V^\sigma). \label{eq:pipj_supp}
\end{align}
With $i=x,y,z$ and $i\neq j$, the coefficients $n_i$ incorporate the in-plane orientation ($n_x=\cos(\phi)\sin(\theta), n_y=\sin(\phi)\sin(\theta)$ and $n_z=\cos(\theta)$) with the azimuthal angle $\phi$ and polar angle $\theta$. The general hopping Hamiltonian reads in momentum space:

\begin{equation}
    \hat{H}(\mathbf{k}) = \sum_{ij} c_i^\dagger(\mathbf{k}) t_{ij}(\mathbf{k}) c_j(\mathbf{k}),
    \label{eq:h_k_supp}
\end{equation}
with the elements $t_{ij}=t_{ji}^*$:
\begin{widetext}
    \begin{align}
        t_{xx}(\mathbf{k}) &= 2V^ \sigma \cos(k_1)+\frac{V^\sigma+3V^\pi}{2}(\cos(k_2)+\cos(k_1+k_2)), \\
        t_{yy}(\mathbf{k}) &= 2V^ \pi \cos(k_1)+\frac{3V^\sigma+V^\pi}{2}(\cos(k_2)+\cos(k_1+k_2)),    \\
        t_{zz}(\mathbf{k}) &= E_z + 2V_{p_z}^\pi (\cos(k_1)+\cos(k_2)+\cos(k_1+k_2)),                              \\
        t_{xy}(\mathbf{k}) &=  -\frac{\sqrt{3}}{2} (V^\pi-V^\sigma)(-\cos(k_2)+\cos(k_1+k_2)),  \\
        t_{xz}(\mathbf{k}) &=  i \lambda_{\rm MIR} \left[2\sin(k_1)-\sin(k_2)+\sin(k_1+k_2)\right], \\
        t_{yz}(\mathbf{k}) &=  \sqrt{3} i \lambda_{\rm MIR} \left[\sin(k_2)+\sin(k_1+k_2)\right].
    \end{align}
\end{widetext}
The integrals $V^\sigma,V^\pi$ and $V_{p_z}^\pi$ denote hoppings within the in-plane subspace and the $p_z$ subspace, respectively. $\lambda_{\rm MIR}$ describes the strength of the mirror symmetry breaking, which couples the in-plane and $p_z$ orbitals. The values can be found in table~\ref{tab:tb_param}.

\subsection*{Atomic SOC}
We consider in our model full $p$-shell atomic spin orbit coupling, which is given in the $\{p_x,p_y,p_z\}$-basis by:
\begin{align}
    \hat{H}^\mathrm{SOC}=&\lambda_\mathrm{SOC} \hat{L} \otimes \hat{S} \\
                  =& \frac{\lambda_{\rm SOC}}{2}\left( \begin{array}{rrr}
                0\hphantom{_x} & -i\sigma_z & i\sigma_y \\
                i\sigma_z & 0\hphantom{_x} & -i\sigma_x  \\
                -i\sigma_y & i\sigma_x & 0\hphantom{_x}   
\end{array}\right).
\end{align}
Its matrix elements can be obtained by explicitly calculating the components of the OAM and spin operator.

\section{ISB on the triangular lattice}
\label{ap_sec:triang_isb}

Here we derive the lattice formulation of the ISB term given in Eq.~\ref{eq:isb_effective} as an effective interaction induced by virtual ISB orbitals i.e. from a substrate, followed by a consideration based on symmetry arguments, only.

In the recently synthesized system of indenene on SiC(0001), the carbon atom of the surface SiC layer located at one of the A/B-points reduces the symmetry of the full lattice to $C_{3\nu}$~\cite{bauernfeind2021design}. To incorporate the hybridization between the triangular lattice and the ISB breaking substrate states in our model, we introduce a virtual $s$-type orbital at site A$=(1/3,2/3)$. This allows us to obtain by down-folding an effective ISB interaction acting on the in-plane orbitals. The Hamiltonian of the $\{p_x,p_y,s\}$-subspace reads:
\begin{align}
    \hat{H}^{ps}=\pmqty{
    \hat{H}^{p_{xy}} & \hat{V}^{p_{xy}s} \\
    \hat{V}^{sp_{xy}} & \hat{H}^{s}
    }.
\end{align}
With the $2\times2$ Hamiltonian of the in-plane subspace $H^{p_{xy}}$, the one-dimensional Hamiltonian of the $s$-subspace $\hat{H}^{s}=E_{s}c_{s}^\dagger c_{s}$, which is given w.r.t. the Dirac-point of the in-plane states, and the hybridization between the two subspaces $\hat{V}^{p_{xy}s}$. The tight-binding elements for an $s$-$p$-overlap $\langle s| H | p_i\rangle = n_iV_{sp}^\sigma$ are given by:
\begin{align}
    t_{sx}(\mathbf{k}) = & \frac{\sqrt{3}}{2}V^{\mathrm{ISB}} \left(e^{\frac{i}{3}(k_1-k_2)}-e^{\frac{i}{3}(-2k_1-k_2)}\right), \\
	t_{sy}(\mathbf{k}) = &V^{\mathrm{ISB}}\left[-\frac{1}{2}\left(e^{\frac{i}{3}(k_1-k_2)}-e^{\frac{i}{3}(-2k_1-k_2)}\right) \right.  \nonumber \\
	                & \hspace{1cm} +\left. e^{\frac{i}{3}(k_1+2k_2)}\right].
\end{align}
Where we substitute directly $V_{sp}^\sigma=V^{\rm ISB}$. By following the lines of~\cite{liu2011low}, an effective low-energy model for the in-plane Dirac states can obtained via downfolding:
\begin{align}
\hat{H}^{eff} \approx \hat{H}^{p_{xy}} -\underbrace{\hat{V}^{p_{xy}s}\cdot (\hat{H}^{s})^{-1} \cdot \hat{V}^{sp_{xy}} }_{\hat{H}^{df}}.
\end{align}
The correction to the in-plane Hamiltonian reads:
\begin{alignat}{3}
t_{xx}^{df}(\mathbf{k})=&\frac{3V_{\mathrm{ISB}}^2}{2E_{s}} &&\left[1-\cos(k_1)\right], \\
t_{yy}^{df}(\mathbf{k})=&\frac{V_{\mathrm{ISB}}^2}{2E_{s}}  &&\left[3+\cos(k_1)-2\cos(k_2) \right. \nonumber \\
             &                                         &&\left.-2\cos\left(k_1+k_2\right)\right],\\
t_{xy}^{df}(\mathbf{k})=&\frac{\sqrt{3}V_{\mathrm{ISB}}^2}{2E_{s}}&& \left[i\{\sin(k_1)+\sin(k_2)-\sin\left(k_1+k_2\right)\}\right. \nonumber \\
             &                                               && \left.+\cos(k_2)-\cos\left(k_1+k_2\right)\right].
             \label{eq:ISB_offd_supp}
\end{alignat}
As $\hat{H}^{p_{xy}}$ vanishes at K/K$^\prime$, the effective Hamiltonian simplifies to the downfolded ISB interaction:
\begin{align}
    \hat{H}^{eff} (K/K^\prime) & = -\frac{9}{4}\frac{V_{\rm ISB}^2}{E_s}\left(\tau_0\mp\tau_y\right).
\end{align}
This gives rise to a rigid band energy shift and promotes orbital angular momentum by $\tau_0$ and $\hat{L}_z=\tau_y$, respectively. As the even terms modify only quantitatively the elements of the full
$p$-basis Hamiltonian (eq. \ref{eq:h_k_supp}), we incorporate the ISB interaction by the effective Hamiltonian given in Eq.~\ref{eq:ISB_rec} with $\lambda_\mathrm{ISB}=-\frac{9}{4}\frac{V_{\rm ISB}^2}{E_s}$.

\begin{figure}
    \centering
    \includegraphics[width=0.7\columnwidth]{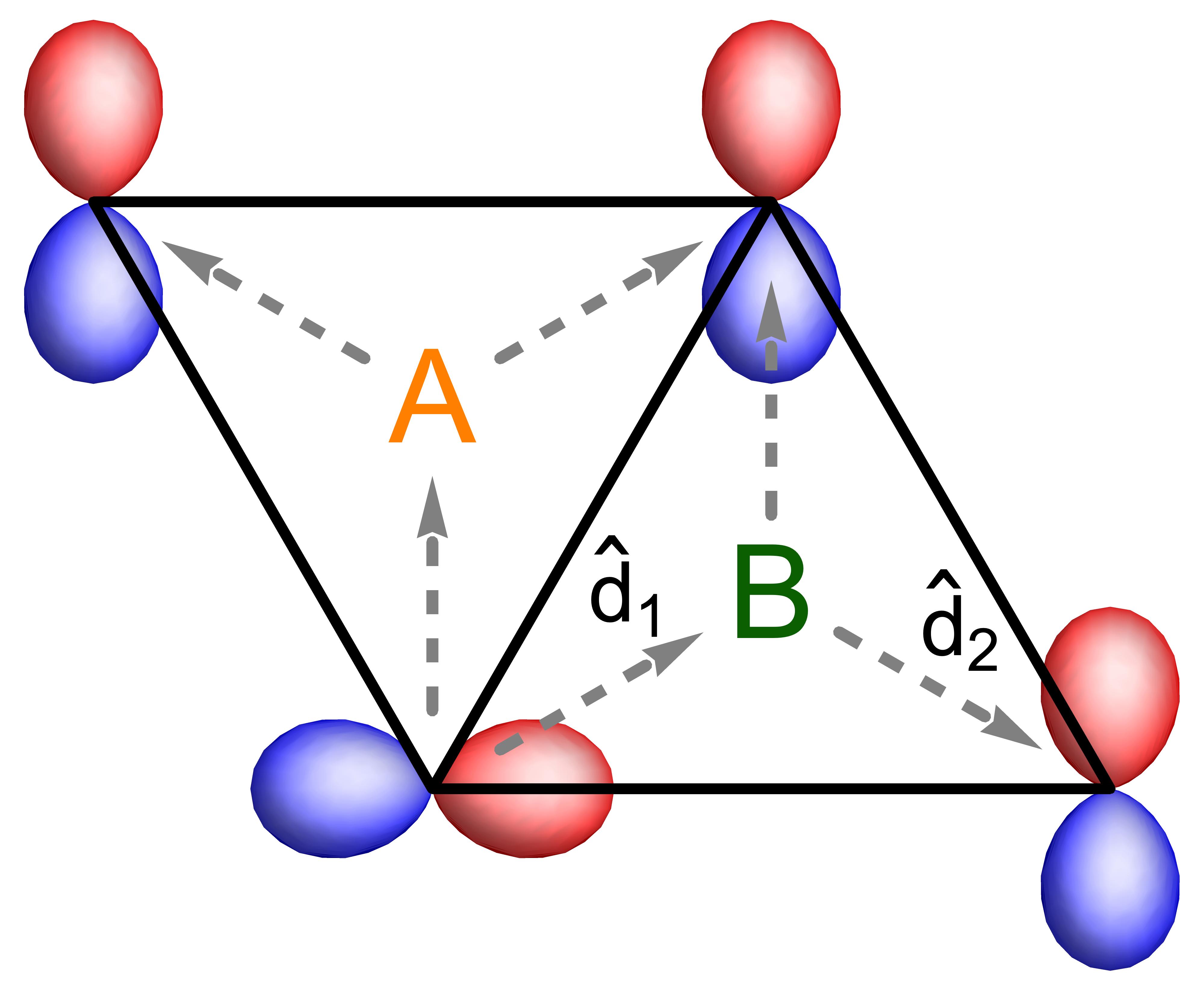}
    \caption{Sketch of the ISB $p_x$-$p_y$ interaction on the triangular lattice activated by a staggered potential peaking at the A/B voids of the triangular lattice. First neighbor hoppings can be decomposed into inequivalent  second order processes via the A/B points as indicated by the gray arrows  $\hat{\mathbf d}_{1}$ and $\hat{\mathbf d}_{2}$.}
    \label{fig:ISB_mech}
\end{figure}

A complementary approach for deriving Eq.~\ref{eq:ISB_real} is to decompose the first-neighbor interaction of the $p_{xy}$ orbitals into a second order hopping processes via the nearest A and B voids as illustrated in Fig.~\ref{fig:ISB_mech}. ISB differentiates the A/B sites and renders the paths inequivalent resulting in different hopping strengths for trajectories through A or B. Further, the $p_x$-$p_y$ hopping process via the A/B sites involves either an overlap of wave functions with same or different parity, which can be defined by the directed angle spanned by the two unit vectors $\hat{\mathbf d}_{1}$ and $\hat{\mathbf d}_{2}$ describing the trajectory. By defining the difference between the hopping strengths through A/B as $\frac{1}{3\sqrt{3}}\lambda_{\mathrm{ISB}}$ i.e., considering only the hoppings through the dominating void site, the interaction can be written as:
\begin{align}
    \hat{H}^\mathrm{ISB} = \frac{\lambda_\mathrm{ISB}}{3\sqrt{3}}\sum_{\langle ij \rangle} \nu_{ij} (c_{p_y,i}^\dagger \sigma_0 c_{p_x,j} +\mathrm{h.c.}). 
    \label{eq:ISB_real_tess}
\end{align}

As a result, an electron travelling from site $i$ to nearest neighbor $j$  experiences a left-right asymmetry, which is reversed when moving from $j$ to $i$. In Eq.~\ref{eq:ISB_real} this is described by the orientation of the Dzyaloshinskii-Moriya vector $\nu_{ij}=(2/\sqrt{3})(\hat{\mathbf d}_1\times \hat{\mathbf d}_2)_z= \pm1$ where $\hat{\mathbf d}_1$ and $\hat{\mathbf d}_2$ are unit vectors pointing from $i$ to A(B) and from A(B) to $j$. This parametrization is reminiscent of that for the second-nearest neighbor SOC term in the Kane-Mele model, which depends on whether the other sublattice appears on the right or on the left of the hopping process~\cite{yao2007spin,kane2005quantum}. However, being not a SOC term but rather involving the orbital degrees of freedom only, this term for the triangular lattice is spin-independent, at odds with the Kane-Mele SOC. A term with a $\mathbf k$-dependence similar to that Eq.~\ref{eq:ISB_rec} but involving second neighbor-processes has been discussed in~\cite{cano2018topology}.

Equation~\ref{eq:ISB_real_tess} is transformed in the spherical harmonics basis by applying the basis transformation $\{p_x,p_y\}\rightarrow\{p_+,p_-\}$
\begin{align}
    \hat{H}^\mathrm{ISB} = \frac{-i}{3\sqrt{3}} \lambda_\mathrm{ISB}\sum_{\langle ij \rangle} \nu_{ij} \sum_{m\in\{\pm |m|\}} m c_{i,m}^\dagger \sigma_0 c_{j,m},
    \label{eq:ISB_real}
\end{align}
and reads in momentum space
\begin{align}
    \hat{H}^{\mathrm{ISB}}(\mathbf{k}) = \frac{2}{3\sqrt{3}}\lambda_{\mathrm{ISB}} &\left[ \sin(k_1)+\sin(k_2)\right. \nonumber\\
     - &\left.\sin\left(k_1+k_2\right)\right] \hat{L}_z\otimes \hat{S}_0.
     \label{eq:ISB_rec}
\end{align}




\section{Sublattice character in the triangular lattice}
We calculate the interference of the Bloch wave function $|\Psi_\mathbf{k}\rangle=\sum_m c_{\mathbf{k},m} e^{i\mathbf{k}\cdot\mathbf{R}} |m\rangle$ at the honeycomb sites ($\mathbf{r}=\{\text{A,B}\}$). By invoking Eq.~\ref{eq:all_shell} and considering only contributions from the $|p_\pm\rangle=|\pm1\rangle$, the momentum-dependent projection weight $X(\mathbf{r})$ is given by:
\begin{align}
    X(\mathbf{r}) &= |\langle \mathbf{r}| \Psi_\mathbf{k}\rangle|^2 \\
    &= \left|\sum_{m\in\{-1,1\}}\frac{1}{3}\sum_{n=1}^3 c_{\mathbf{k},m} e^{i\mathbf{k}\cdot\mathbf{R}}\langle \mathbf{r}|m_{\mathbf{R}_n}\rangle\right|^2 \\
    &= \left|\sum_{m\in\{-1,1\}}\frac{1}{3}\sum_{n=1}^3 c_{\mathbf{k},m} e^{i\left(\mathbf{k}\cdot\mathbf{R}+\frac{2\pi}{3}nm\right)}\right|^2.
    \label{eq:Bloch_loc_weight}
\end{align}

\section{Construction of the overlap matrix}
Here we follow the recipe of Ref.~\cite{Soluyanov2011wannier}. The calculation is straight-forward if the trial orbitals and the cell-periodic Bloch wave functions are spanned by the same basis set. For our model, this is the case for the triangular $p$-shell $J_{1/2}$-trial basis. The projection onto trial orbitals on the honeycomb sites requires however a momentum-dependent basis transformation as the Bloch and the trial basis have different Wyckoff positions, which results in a Bloch phase difference.

In analogy to Eq.~\ref{eq:ShiWul_proj}, we define the overlap of the Bloch wave function of the triangular $p$-model ($m=\{-1,0,1\}$) with the honeycomb trial basis $\ket{\tau_j}$ at position $\mathbf{x}_j$ as
\begin{align}
    \braket{\Psi_{n\mathbf{k}}|\tau_j} = \sum_{\mathbf{R}_j} e^{i\mathbf{k}\cdot(\mathbf{x}_j-\mathbf{R}_j)}\sum_m c_m^* \braket{m|\tau_j}.
\end{align}
Without loss of generality, we consider only the contribution of the three nearest-neighbor sites $\mathbf{R}_j$ to the Bloch wave function, which are for the A(B) site $\mathbf{R}_j=\{(0,0),(0,1),(1,1)\}$ ($\mathbf{R}_j=\{(0,0),(1,0),(1,1)\}$). Further, also the parity w.r.t. $z\mapsto-z$ has to be considered: the triangular $m=\pm1$ in-plane orbitals map onto $s$ orbitals on the honeycomb site, while a $p_z$ ($m=0$) orbital on the triangular Wyckoff position is mapped onto a $p_z$ orbital on the honeycomb sites.

\section{Local OAM winding promoted real-space obstruction}
\label{ap_sec:obstruction}

\begin{figure}
    \centering
    \includegraphics[width=\columnwidth]{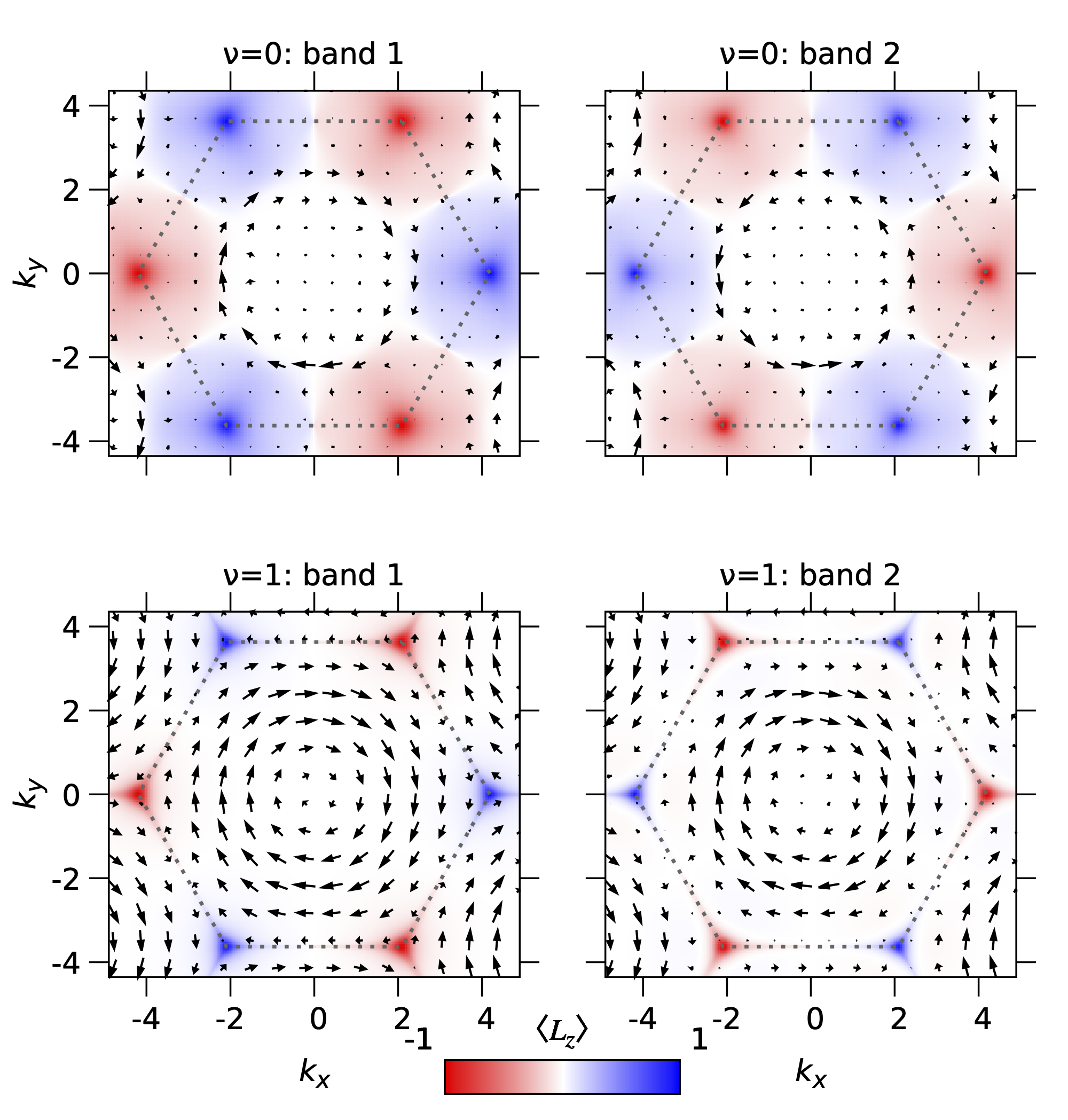}
    \caption{Local orbital angular momentum polarization of the two valence bands in the $\nu=0$ and $\nu=1$ phases. To lift the degeneracy of the valence bands at the valley momenta, a small values of $\lambda_\text{ISB}=0.1$ has been chosen and in the $\nu=0$ the compensating in-plane OAM is visualized by setting $\lambda_\text{MIR}=0.01$ (otherwise the in-plane OAM vanishes). The dashed lines indicate the first BZ.}
    \label{fig:L_map}
\end{figure}

Based on the wave function symmetry of the valence bands, we will argue in the following, that a non-obstructed Wannier basis cannot exist in the QSHI phase. We assume that the bands are energetically sufficiently isolated, to allow for a single band description.
As illustrated in Fig.~\ref{fig:model_no_isb}, in the absence of SOC, the valence band is $p_z$-type at $\Gamma$, has a $p_\pm$ degeneracy at K and has radial (w.r.t. the nearest $\Gamma$ point) in-plane character at the three M points ($\ket{p_r}\propto\alpha \ket{p_x}+\beta \ket{p_y}$). This results in a metallic band crossing of the two lowest bands which gives rise to a nodal line (see red-green band crossing in Fig.~\ref{fig:model_no_isb}).
In the presence of horizontal mirror symmetry breaking the nodal ring gaps-out as the radial in-plane and the $p_z$ band hybridize by forming tangential angular momentum states of the form $\ket{m_{tan}}=1/\sqrt{2}(\ket{p_z}\pm i \ket{p_r}$. Upon considering SOC, a non-trivial gap is opened and the valence states at the K/K$^\prime$ read: $\ket{l,m_z}=\ket{\pm1,\mp1/2}$ states. If considering the whole BZ, the local orbital angular polarization of the valence bands will cover the whole unit sphere as shown in Fig.~\ref{fig:L_map}. For a trial projection basis located on the triangular site, there must be momenta, where at least one trial orbital is orthogonal to both valence states which prohibits a non-obstructed Wannier construction. 

\begin{figure}
    \centering
    \includegraphics[width=\columnwidth]{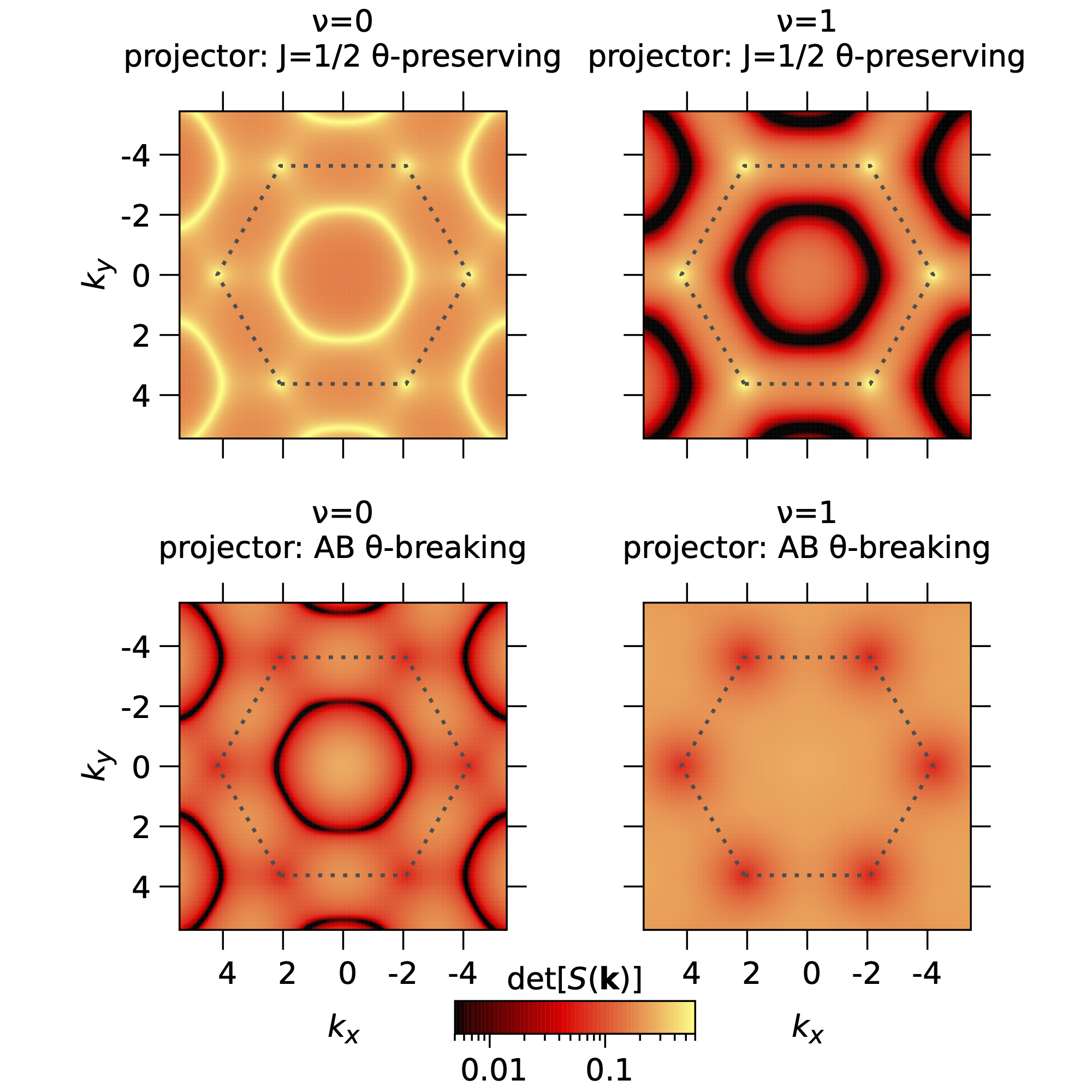}
    \caption{Logarithmic plot of det[$S(\mathbf{k})$] of the vertical-reflection symmetric model $\lambda_\text{ISB}=0$ for a time-reversal-symmetric $J=1/2$ and time-reversal-asymmetric A/B trial basis sets. The dashed lines indicate the first BZ.}
    \label{fig:det_S_map}
\end{figure}

The importance of the nodal line is further supported by inspecting det[$S(\mathbf{k})$] in the whole BZ as shown in Fig.~\ref{fig:det_S_map}. The overlap eigenvalues of the $J_{1/2}$ trial basis are largest in the trivial phase at the momenta of the SOC gapped nodal line and at K/K$^\prime$. In the non-trivial phase, the vanishing eigenvalues occur along the nodal ring. In turn, the time-reversal symmetry breaking trial basis on the honeycomb sites introduced in Eqs.~\ref{eq:Sol_AB_proj} and~\ref{eq:spz_hyb_proj} has a finite overlap in the $\nu=1$ phase, but vanishing overlap along the momenta of the nodal line in the $\nu=0$ phase.

\section{GW methods}
\label{ap_sec:GW}

\begin{figure}
    \centering
    \includegraphics[width=1.0\columnwidth, trim={4.2cm 3.2cm 2cm  0cm},clip]{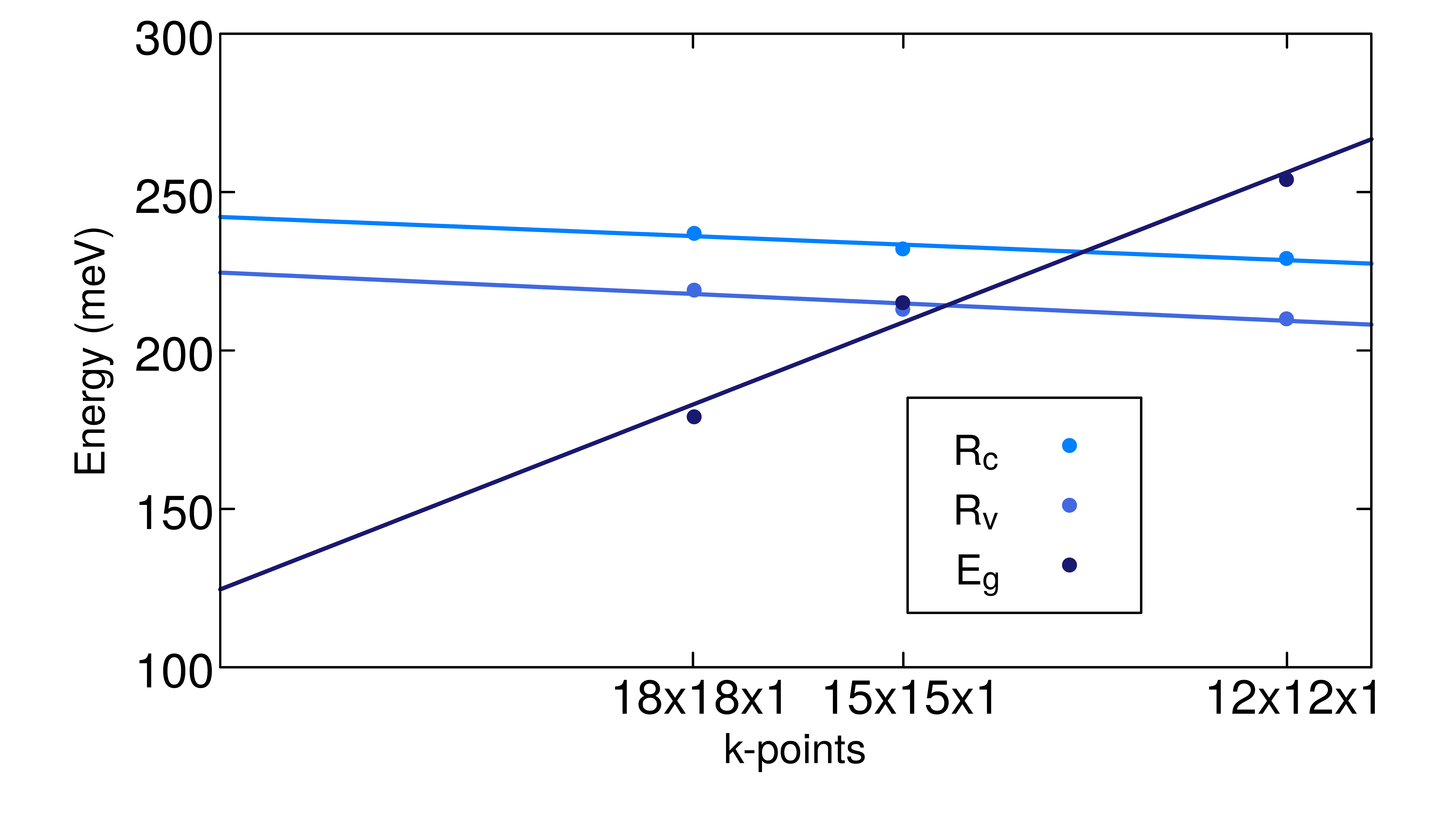}
    \caption{$GW$ scaling of the energy gap ($\mathrm{E_g}$), Rashba splitting in the conduction band ($\mathrm{R_c}$) and Rashba splitting in the valence band ($\mathrm{R_v}$), as a function of the $k$-points, in the energy range [100: 300] meV. The dots correspond to the computed values and the solid lines to the interpolation of the results.}
    \label{fig:scaling}
\end{figure}

Our first-principles study started defining, via ionic relaxation, the equilibrium geometry of the system; this is composed by an indenene layer and four substrate layers. A big enough number of substrate SiC layers has been considered in order to correctly determine the screening properties and the low-energy physics.  Its importance becomes indeed evident, especially for $GW$ calculations, when the number of layers is changed. When this number is increased, increasing accordingly also the unit-cell height to preserve the same amount of vacuum, the energy gap decreases, because of the increased dielectric screening.\\
Also, when a fixed number of substrate layers is considered, rather increasing the amount of vacuum along $z$, it is found an increasing of the energy gap value.\\
Both self-consistent (\emph{scf}) and non-self-consistent (\emph{nscf}) DFT  calculations are based on the projector augmented wave (PAW) method and the generalized gradient approximation (GGA) within the Perdew-Burke-Ernzerhof (PBE) scheme~\cite{PhysRevLett.77.3865}, as implemented in the Vienna Ab-initio Simulation Package (VASP)~\cite{PhysRevB.59.1758,PhysRevB.54.11169}. A plane-wave cutoff of 500 eV has been used, together with a $15\times 15 \times 1$ $k$-mesh.\\ 
SOC has been included self-consistently and, because of the small value of the energy gap ($\sim 50$ meV, in DFT), the width of the
Gaussian smearing has been chosen equal to 0.001 eV; this allowed us to obtain a sharp transition from the occupied states to the unoccupied ones.\\
Subsequent single-shot $GW$ calculations show an almost independent behavior with respect to the number of unoccupied states starting from a reasonably high threshold; we considered then 338 empty bands. For the same independence's reasons, we set the energy cut-off for the response function to 50 eV.\\ 
The number of imaginary time grid points has been chosen equal to 100, also because this parameter does not influence in a meaningful way the computation time.\\
While the value of the Rashba splitting in the valence and conduction bands resulted to be always almost independent from the meaningful parameters, at least for a not-too-drastic change of them, the analysis of the energy gap required more cautions.\\
We choose an optimal setup provided by a unit-cell of 45 \AA $~$along the $z$-direction and four layers of substrate.\\ 
The last check is given by the analysis of the low-energy electronic structure as a function of $k$-points number. Only in this case the value of the Rashba splittings exhibits a slightly dependence, while the energy gap exhibits a strong one (see Fig.~\ref{fig:scaling}). The interpolated values are ideally obtained from an infinitely dense $k$-mesh, and they have been used to realize the correct $GW$ band structure plot (see Fig.~\ref{fig:GW_exp}), subsequently to the wannierization procedure~\cite{MOSTOFI20142309}.
\section{ARPES}
\label{ap_sec:ARPES}
ARPES measurements were acquired in our home-lab setup equipped with a
hemispherical analyzer (PHOIBOS 100), an unmonochromatized He-VUV lamp (UVS 300, 21.2$\,$eV) light source, and a 6-axis manipulator capable of
LHe-cooling to $20\,\mathrm{K}$. Differential pumping of the He-VUV-lamp kept the
base pressure below $1\cdot10^{-10}\,$mbar during the ARPES measurements.

\section{STM and STS}
\label{ap_sec:STM_STS}
STM and STS measurements were taken with a commercial Omicron LT-STM operated at $4.7\,\mathrm{K}$ and a base pressure lower than $5\cdot10^{-11}\,$mbar. Before and after each measurement the chemically etched W-tip was prepared on an Ag(111) single-crystal. Tunneling parameter of $\dd I/\dd V$-curves and images shown in Fig.~\ref{fig:GW_exp} are summarized in Tab.~\ref{Table_STM}. 
The STS measurement presented in Fig.~\ref{fig:GW_exp}e was recorded with a standard lock-in technique  (modulation voltage of V$_{\text{rms}}= 10\,$mV and modulation frequency of 971$\,$Hz). 

\setlength{\tabcolsep}{9pt}
\begin{table}
\begin{tabular}{|c|c|c|c|} 
\hline
Fig. & $U_{\mathrm{Bias}}\,$ & $I_\mathrm{T}$ & tip approach \\
\hline
5 (d) & 2.0$\,$V & 250$\,$pA & 0$\,$nm (z$_0$) \\ 
5 (d) (inset) & -0.85$\,$V & 50$\,$pA & -0.24$\,$nm (z$_1$)\\ 
5 (e) & -0.9$\,$V & 50$\,$pA & -0.28$\,$nm \\
\hline
\end{tabular}
\caption{Bias voltage $U_{\mathrm{Bias}}\,$ and tunneling current $I_\mathrm{T}$ stabilization parameters for STS shown in Fig.~\ref{fig:GW_exp}. In some STS measurements the tip has been approached after the feedback loop was switched off. Note that the tip sample distance is also affected by the tunneling setpoint itself, i.e. $U_{\mathrm{Bias}}\,$ and $I_\mathrm{T}$. In the case of panel Fig.~\ref{fig:GW_exp}(d) this yields an additional height difference of 5.4$\mathrm{\AA}$, and thus a total height difference between z$_0$ and z$_1$ of z$_1$=z$_0$-7.8$\, \mathrm{\AA}$.}
\label{Table_STM}
\end{table}

\FloatBarrier

\bibliographystyle{apsrev4-2_prx.bst}
\bibliography{prb.bib}

\end{document}